# The three kingdoms – Photoinduced electron transfer cascades controlled by electronic couplings

Guangjun Yang,[a] Georgina E. Shillito,[a] Clara Zens,[a] Benjamin Dietzek-Ivanšić[a,b] and Stephan Kupfer[a]*

Dedicated to Prof. Wolfgang Weigand on occasion of his 65[th] birthday.

[a]  G. Yang, Clara Zens, Dr. G. E. Shillito, Dr. S. Kupfer, Prof. Dr. B. Dietzek-Ivanšić

Institute of Physical Chemistry, Friedrich Schiller University Jena,

Helmholtzweg 4, 07743 Jena, Germany

E-mail: stephan.kupfer@uni-jena.de

[b]  Prof. Dr. B. Dietzek-Ivanšić

Leibniz Institute of Photonic Technology (IPHT) e.V. Department Functional Interfaces,

Albert-Einstein-Straße 9, 07745 Jena, Germany

**Abstract**

Excited states are the key species in photocatalysis, while the critical parameters that govern the potential applications of such excited states are their: i) excitation energy, ii) accessibility, and iii) lifetime. However, in molecular transition metal-based photosensitizers there is a design tension between the creation of long-lived excited (triplet), *e.g.*, metal-to-ligand charge transfer ($^3$MLCT) states and the population of such states. Long-lived triplet states have low spin-orbit coupling (SOC) and hence their population, either by direct photoexcitation or via subsequent excited state relaxation is low. Thus, a long-lived triplet state can be populated but inefficiently. If the SOC is increased, the triplet state population efficiency is improved – coming at the cost of decreasing the lifetime. A promising strategy to isolate the triplet excited state away from the metal after intersystem crossing (ISC) involves the combination of the transition metal complex and an organic donor/acceptor group. Here we elucidate the excited-state branching processes in a series of Ru(II)-terpyridyl push-pull triads by means of quantum chemical simulations. Scalar-relativistic time-dependent density theory simulations reveal that efficient ISC takes place along $^{1/3}$MLCT-gateway states. In the following, competitive electron transfer pathways involving the organic chromophore, *i.e.*, 10-methylphenothiazinyl and the terpyridyl ligands are available. The kinetics of the underlying electron transfer processes were investigated within the semi-classical Marcus picture. The electron transfer kinetics were described along efficient internal reaction coordinates that connect the respective photoredox intermediates. The key parameter that governs the population transfer away from the metal either towards the organic chromophore either by means of ligand-to-ligand ($^3$LLCT; weakly coupled) or intra-ligand charge transfer ($^3$ILCT; strongly coupled) states was determined to be the magnitude of the involved electronic coupling.

# 1 Introduction

Solar energy conversion is among the most promising approaches to transform our energy sector towards sustainability.[1-10] In this context, supramolecular photocatalysis allows the conversion of sunlight into chemical energy such as molecular hydrogen as well as the conversion of *e.g.*, carbon dioxide into commodity chemicals such as formaldehyde or methanol.[11-17] In such photocatalytic processes, molecular excited states are the key species, with the critical parameters such as excitation energy, accessibility and lifetime governing their potential applications. The character of the excited state also plays a prominent role, with charge-separated excited states being of particular relevance to photocatalytic applications. In order to meet these criteria, 4d and 5d transition metal complexes are most often used as photosensitizers due to their favourable photophysical and electrochemical properties, alongside thermal, light and pH stability.[18-21] Although considerable progress has been made in the application of earth abundant 3d metal-based photocentres, these systems still suffer from rather short excited state lifetimes due to the presence of ultrafast deactivation pathways. Therefore, in molecular transition metal-based photosensitizers there is a design tension between the creation of long-lived excited (triplet) states and the population of such states. Long-lived triplet states have low spin-orbit coupling (SOC) and hence their population, either by direct photoexcitation or via subsequent excited state relaxation is low. Thus, a long-lived triplet state can be populated but inefficiently. If the SOC is increased, the triplet state population efficiency is improved – however, this is accompanied by a decrease in the excited state lifetime.[22, 23] In this regard, the efficiency of intersystem crossing (ISC) and in consequence the population of charge-separated triplet states in transition metal complexes is highly beneficial. In particular, metal-to-ligand charge transfer ($^1$MLCT) excitations involved in the light-harvesting processes typically provide pronounced SOCs and allow an efficient population transfer to potentially long-lived charge-separated states, *e.g.*, of $^3$MLCT character. However, at the same time, the SOCs between the (emissive) $^3$MLCT states and the singlet ground state foster radiative deactivation processes, while even stronger SOCs among low-lying $^3$MC (metal-centred) states and the singlet ground state may yield a rapid deactivation of the charge-separated state, particularly in the case of 3d metal-based compounds.[24-28] An alternative to transition metal complexes in the frame of light-harvesting molecules, are organic chromophores. However, long-lived charge-separated triplet states in organic dyes are often inaccessible or populated inefficiently due to small SOCs.

One promising molecular approach to enable panchromatic absorption in the visible spectral region as well as an efficient population of long-lived charge-separated states is to combine organic and inorganic chromophores into one structure. This strategy allows light-harvesting by both chromophores, efficient ISC along a $^{1/3}$MLCT gateway and subsequently isolating the triplet excited state away from the metal – onto the organic chromophore, *e.g.*, by means of intraligand or ligand-to-ligand charge transfer ($^3$ILCT and $^3$LLCT) states, see Figure 1a), or by virtue of energy transfer channels.[29-32] Therefore, pronounced lifetimes of the charge-separated 'trap state' (*e.g.*, $^3$ILCT and $^3$LLCT) can be realized as these $^3\pi\pi^*$ states are only weakly coupled to the singlet ground state.

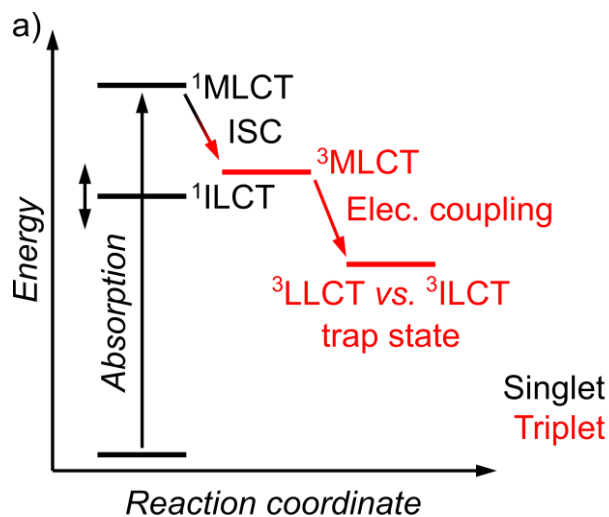

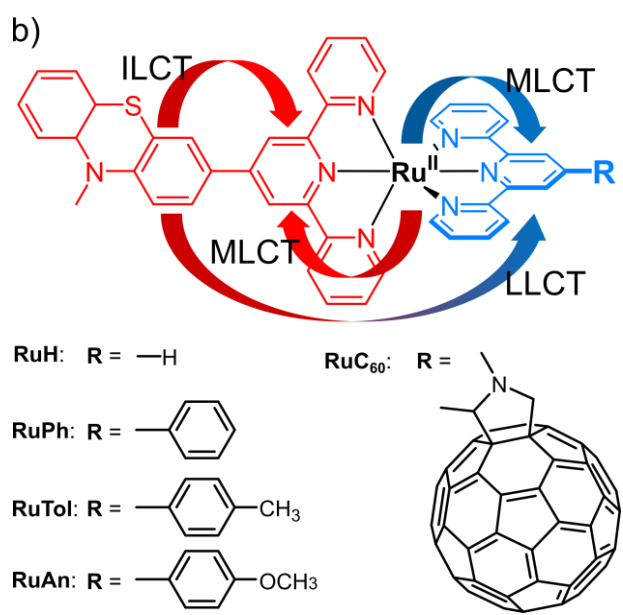

**Figure 1.** a) Jablonski scheme visualizing excited state processes associated with intersystem crossing and the population of long-lived 'trap states' by means of the associated electronic coupling. b) Structure of investigated ruthenium(II) photosensitizers (**RuR**) incorporating a phenothiazine-tpy donor ligand (red) as well as a tpy-based acceptor ligand (blue). Prominent electronic transitions are indicated.

In previous joint synthetic-spectroscopic-theoretical investigations, we focused on unravelling the Franck-Condon photophysics of such dyads which combined inorganic and organic chromophores as well as on the description of ISC processes along $^{1/3}$MLCT gateway states for a series of triphenylamine-donor-ligand Rh(I) and Pt(II) complexes[33-35] and a Ru(II)-based dye incorporating a thiazole push-push ligand.[36] The nature of the $^{1/3}$MLCT-gateway was tuned systematically by structural modification at the metal-centre[33], the triphenylamine-donor-ligand[34] as well as by solvent effects[35]. Consequently, remote control of excited state relaxation channels within the triplet manifold was rationalized, leading either to short-lived $^3$MLCT or long-lived $^3$ILCT states.

In the scope of the present contribution, we aim to elucidate the light-driven processes and excited state relaxation cascades associated with the population of such charge-separated triplet states localized away from the metal centre in a series of Ru(II)-terpyridyl push-pull triads by means of quantum chemical simulations. Along this series of dyes, one terpyridyl (tpy) ligand is substituted with an electron donating 10-methylphenothiazinyl (PTZ) moiety, while the substitution pattern of the second tpy ligand is systematically modified from the parent ligand architecture (R = H; **RuH**) by a phenyl (**RuPh**), tolyl (**RuTol**), anisyl (**RuAn**) or by a $C_{60}$ fullerene (**RuC$_{60}$**), see Figure 1b). The effect of the substitution pattern is particularly interesting as previous experimental studies highlighted the potential to remotely control the photo-induced electron transfer kinetics by means of structural modification. Based on ns-transient absorption spectroscopy and electrochemistry it was shown that the photoinduced electron transfer (ET) kinetics are mainly modulated by the electronic coupling ($V_{DA}$) between the donor (D) and acceptor (A) states and to a lesser extent by the underlying thermodynamic properties such as the driving force ($\Delta G$) and reorganization energy ($\lambda$).[37]

In this study, quantum chemical methods, *i.e.*, density and time-dependent density functional theory (DFT and TDDFT) as well as multiconfigurational simulations, are utilized to investigate the light-driven processes for the series of **RuR** push-pull triads. Prominent excited state relaxation channels involved in the population transfer from the singlet to the triplet manifold are investigated by means of scalar-relativistic TDDFT (SR-TDDFT), while the focus of the present computational study is set to address the substitution effect on the subsequent ET kinetics among key $^3$MLCT, $^3$LLCT and $^3$ILCT states based on semi-classical Marcus theory. Thereby, we follow our recently introduced protocol to assess the kinetics of intramolecular ET processes along efficient reaction coordinates within the Marcus picture as benchmarked with respect to (dissipative) quantum dynamics.[38-40]

## 2 Computational Details

All quantum chemical calculations, if not stated otherwise, were performed utilizing the Gaussian 16 program[41]. The singlet ground state equilibrium structures and electronic properties of the Ru(II) complexes as ruthenium(II)-based triads, *i.e.* **RuH**, **RuPh**, **RuTol**, **RuAn**, **RuC$_{60}$**, were obtained at the density functional level of theory (DFT) using the B3LYP[33, 34, 42-45] exchange correlation (XC) functional in association with the def2-SVP[46, 47] basis set as well as the respective core potentials. Subsequently, a vibrational analysis was carried out for each optimized ground state structure to verify that a (local) minimum on the 3*N*-6 dimensional potential energy (hyper-)surface was obtained. The effects of interaction with the dichloromethane solvent (CH$_2$Cl$_2$: $\varepsilon$ = 8.93, *n* = 1.4070) were taken into account by the solute electron density (SMD) variant of the integral equation formalism of the polarizable continuum model (IEFPCM).[48, 49] All calculations were performed including D3 dispersion correction with Becke-Johnson damping (D3BJ).[50]

Thereafter, time dependent DFT (TDDFT) calculations were performed for these five complexes using the same XC functional and basis set as mentioned above in the preceding ground state calculations. These TDDFT simulations aim to provide insight into the exited state properties, *i.e.* excitation energies, oscillator strengths, transition dipole moments and electronic characters of the 100 lowest excited singlet states as well as of the 20 lowest triplet states within the Franck-Condon region. Solvent effects on the Franck-Condon photophysics, where only the fast reorganization of the solvent is important, were addressed by the non-equilibrium procedure of solvation. This computational protocol was already successfully applied to elucidate the ground and excited-state properties of structurally closely related Ru(II) complexes and allows a balanced description of metal-to-ligand charge transfer (MLCT), intra-ligand charge transfer (ILCT), ligand-to-ligand charge transfer (LLCT) as well as local intra-ligand states.[38, 51-58] Furthermore, to confirm the order of the low-lying singlet excited-states in the Franck-Condon point, the 50 lowest energy singlet transitions of **RuH** were evaluated using the double-hybrid SOS-wPBEPP86 functional [59] – also in association with as the def2-SVP basis set and respective core potentials as implemented in Orca 5.0. Solvent effects were included using the conductor-like polarizable continuum model (CPCM)[60] for CH$_2$Cl$_2$.

While dynamic correlation, *i.e.*, the close-range interaction between neighboring electrons, is typically well described by (TD)DFT, static correlation stemming from near-degenerate electronic configurations (Slater determinants) is insufficiently treated by (TD)DFT methods due to its single-determinant nature. To account for static correlation, multiconfigurational methods, *e.g.*, the complete active space self-consistent field (CASSCF) approach,[61] are the methods of choice. The application of multiconfigurational simulations provides an unbiased description of the photophysics of small to medium sized chemical systems, *e.g.*, the light-driven charge transfer processes for the given set of Ru(II)-based triads. Yet, the computational demand quickly increases with the size of the active space (AS). In case of **RuH** (smallest triad), an appropriate active space (AS) would include a Ru-centered AS comprising ten electrons in seven molecular orbitals, (10,7), a (18,18) with the $\pi_{\text{tpy}}/\pi^*_{\text{tpy}}$ system of each terpyridine (tpy) ligand, as well as a (16,14) containing the $\pi_{\text{PTZ}}/\pi^*_{\text{PTZ}}$ orbitals of the 10-methylphenothiazinyl moiety (including sulphur's and nitrogen's lone pairs in the aromatic plane). Consequentially, an AS (62,57) is obtained for **RuH**, which is unfeasible without further restrictions. To restrict the number of configuration state functions (CSFs) in the CASSCF methodology, several approaches have been introduced, *e.g.*, the restricted active space (RAS)SCF[62-64] method, which allows the computational demand to be reduced by splitting the AS into three subspaces. RAS1 holds mostly doubly occupied orbitals with a predefined number of maximal electron holes, while the RAS3 subspace contains mostly unoccupied orbitals where a defined number of electrons are allowed to be excited into. Finally, a full configuration interaction calculation is performed within RAS2 – equivalent to the AS in the CASSCF method. To label the RAS calculations, the notation RAS ($n,l,m;i,j,k$)[65] is used. Here, the index $n$ indicates the number of active electrons, $l$ label the maximum number of holes in the RAS1, and $m$ is the maximum number of electrons in the RAS3. The labels $i$, $j$, and $k$ refer to the number of active orbitals in the RAS1, RAS2, and RAS3 subspaces, respectively. All RASSCF calculations were performed as implemented in OpenMolcas 22.02[66, 67] using the singlet ground state geometry of **RuH** obtained at the B3LYP level of theory. The 6-31G(d) double-ζ basis set[68] as well as the MWB28[69] relativistic core potential were applied.

The RAS partitioning of **RuH** was built based on our experience with a structurally related push-pull Ru(II)-based polypyridyl dye which combines inorganic and organic chromophores.[36] In order to provide a proper description of MLCT, ILCT and LLCT states of interest, the RAS comprises the (10,7) of the ruthenium atom which includes two pairs of $\sigma/\sigma^*$ orbitals and the three $t_{2g}$ orbitals ($d_{xy}$, $d_{xz}$ and $d_{yz}$), four pairs of $\pi_{tpy}/\pi^*_{tpy}$ orbitals (four orbitals per tpy ligand), three pairs of $\pi_{PTZ}/\pi^*_{PTZ}$ orbitals as well as one non-bonding orbital of the chelating nitrogen atoms which showed pronounced mixing σ orbitals of the coordination environment. Thereby, the RAS2 contains the three $t_{2g}$ orbitals, the highest occupied molecular orbital of PTZ group and the lowest unoccupied molecular orbital of each tpy ligand. The remaining either occupied or unoccupied orbitals within the Hartree-Fock reference wavefunction were distributed over the RAS1 and RAS3 subspaces, accordingly. In consequence, a RAS (26,2,2;9,6,7) is obtained, see Figure 5, which spans over almost 700,000 and more than 1.2 million CSFs in singlet and triplet multiplicity, respectively. State-average (SA-)RASSCF calculations were carried out considering the first nine singlet and triplet roots, respectively. The transition dipole moments were obtained at the SA-RASSCF level of theory using the CAS state interaction method exclusively for the singlet roots.[70]

To assess scalar-relativistic effects and their potential impact on the subsequent excited-state relaxation pathways accessible upon intersystem crossing (ISC) within the Franck-Condon region, scalar-relativistic TDDFT calculations were performed utilizing Orca 5.0 using the scalar-relativistic zeroth-order regular approximation (SR-ZORA). DFT and TDDFT calculations were performed using the B3LYP ("Gaussian version") XC functional;[71] the SARC-ZORA-TZVP[72] basis set was utilized for ruthenium, while all other atoms were described using the respective def2-TZVP basis sets (with the corresponding SARC/J auxiliary basis set).[73] The 20 lowest singlet-singlet and singlet-triplet excitations were obtained, while spin-orbit couplings (SOCs) between these states and the singlet ground state were obtained at the SR-ZORA-TDDFT level of theory. Theeffects of interaction with $CH_2Cl_2$ were taken into consideration at the CPCM level of theory.

Furthermore, equilibrium geometries of three specific excited-states involved in the subsequent excited-state relaxation cascade within the triplet manifold, *i.e.* of the low-lying ³MLCT as well as of the ³LLCT and ³ILCT states, were optimized for all ruthenium complexes.

These states were fully relaxed at the TDDFT level of theory using our external optimizer pysisyphus[74] – interfaced in the present case, with Gaussian 16 for gradient and energy calculations. Wavefunction overlaps[75] were utilized to track excited-state characters (*i.e.* $^3$MLCT, $^3$LLCT and $^3$ILCT) along the course of the optimization. The equilibrium procedure of solvation SMD was applied for all optimizations.

To access photoinduced electron transfer (ET) processes in Ru(II)-bridged donor-acceptor dyads, semi-classical Marcus theory was applied. According to Marcus theory, ET processes occur along the parabolic diabatic potential energy curves (PECs) of the electron donor state (D; *i.e.* $^3$MLCT) and the acceptor state (A; *i.e.* $^3$LLCT vs. $^3$ILCT) along the reaction coordinate $R_{ET}$. Structural distortion within the donor state – induced by thermal fluctuations of the surrounding bath (*e.g.*, the solvent) – may provide sufficient electronic coupling between D and A to yield a population transfer between the electronic states of interest. Herein, the rate constant, $k_{ET}$, for such an ET is given within the Marcus-picture by:

$$k_{ET} = \frac{2\pi}{\hbar}|V_{DA}|^2(4\pi\lambda k_B T)^{-\frac{1}{2}}\exp\left(-\frac{(\Delta G+\lambda)^2}{4\lambda k_B T}\right) \quad \text{Equation (1)}$$

Where $V_{DA}$ denotes electronic coupling between the D and A states at the crossing point of the diabatic PECs, $\lambda$ is the reorganization energy, $\Delta G$ represents the driving force, *i.e.* the Gibbs free energy, for the ET reaction, $k_B$ is Boltzmann constant, $T$ is absolute temperature.

In case of all present complexes, D and A states of interest are of triplet multiplicity. The ET kinetics for the different pairs of D/A states were described along a linear-interpolated internal coordinate (LIIC) connecting the optimized equilibrium structures of the donor and acceptor states – as obtained by pysisyphus. All optimized and interpolated structures are available from the online repository Zenodo via Ref. [76]. The diabatic PECs for D and A were constructed along the LIIC (denoted $R_{ET}$) by means of TDDFT single-point calculations. The electronic coupling $V_{DA}$ between the $^3$MLCT (D) and $^3$LLCT (A) as well as between the $^3$MLCT (D) and $^3$ILCT (A) states, were obtained by a unitary transformation of the adiabatic states, $V_1^{ad}(R_{ET})$ and $V_2^{ad}(R_{ET})$, to the respective diabatic states, ($V_D(R_{ET})$ and $V_A(R_{ET})$):

$$\begin{pmatrix} V_A(R_{ET}) & V_{AD}(R_{ET}) \\ V_{DA}(R_{ET}) & V_D(R_{ET}) \end{pmatrix} = U^{-1}\begin{pmatrix} V_1^{ad}(R_{ET}) & 0 \\ 0 & V_2^{ad}(R_{ET}) \end{pmatrix}U \quad , \quad \text{Equation (2)}$$

in which U is a general Unitary matrix, *i.e.,* $\begin{pmatrix} \cos\theta & \sin\theta \\ -\sin\theta & \cos\theta \end{pmatrix}$. The electronic coupling is then given by:

$$V_{DA} = \frac{1}{2}|V_2^{ad} - V_1^{ad}|\sin(2\theta) \quad , \qquad \text{Equation (3)}$$

while the dependency on mixing angle *θ* is expressed within the generalized Mulliken–Hush (GMH) method by means of the permanent and transition dipole moments of the involved states:

$$\sin(2\theta) = \frac{2\mu_{12}}{\sqrt{|\mu_{11}-\mu_{22}|^2 + 4|\mu_{12}|^2}} \quad , \qquad \text{Equation (4)}$$

or using the fragment charge difference (FCD) approach:

$$\sin(2\theta) = \frac{2\overline{\Delta q}_{12}}{\sqrt{|\Delta q_{11}-\Delta q_{22}|^2 + 4|\overline{\Delta q}_{12}|^2}} \qquad \text{Equation (5)}$$

Finally, $V_{DA}$ is defined at the crossing point (*θ* = 45°) of the diabatic states by the minimum splitting method:

$$V_{DA} = \frac{1}{2}|V_2^{ad} - V_1^{ad}|_{min} \quad . \qquad \text{Equation (6)}$$

This computational approach was lately introduced in the scope of photoinduced intramolecular ET processes in photocatalysis[38, 58, 77] as well as to assess the competitive energy and electron transfer processes in light-harvesting antennae.[78]

Additionally, the electronic couplings were obtained based on the GMH method as well as with the fragment charge difference (FCD) approach, which are widely applied to study inter- and intramolecular electron transfer processes.[79-84] These simulations were performed at the B3LYP/def2-SVP level of theory using Q-Chem.[85] Solvent effects (CH$_2$Cl$_2$) were taken into account using a the conductor-like polarizable continuum model.

## 3   Results and Discussion

The following section addresses the light-driven processes for the present set of Ru(II)-based triads as predicted at the (scalar-relativistic) time-dependent density functional level of theory (SR-TDDFT). Initially, structural and electronic properties within the Franck-Condon point as well as the nature of the electronic transitions underlying the UV-vis absorption bands are carefully evaluated and compared with respect to the substitution pattern. Furthermore, multiconfigurational simulations – based the restricted active space self-consistent field methodology – were utilized to benchmark the cost-efficient TDDFT simulations. Subsequently, the population transfer channels among the optical accessible excited singlet states, and energetically close-lying triplet states are identified. Finally, the kinetics of the photoinduced electron transfer processes, populating the (long-lived) charge-separated species, are simulated within the semi-classical Marcus picture. These simulations assess the charge separation efficiency of excited states governed by the thermodynamical properties and the electronic coupling of the involved donor and acceptor states.

## 3.1 Frank-Condon Photophysics

Initially, the Franck-Condon photophysics of the Ru(II) complex **RuH**, Figure 1b), were carefully evaluated using TDDFT as well as by means of multiconfigurational simulations. Figure 2 shows the experimental electronic absorption spectrum in $CH_2Cl_2$ alongside the simulated spectra obtained by spin-free (SF-)TDDFT as well as by means of SR-TDDFT using the B3LYP hybrid functional.

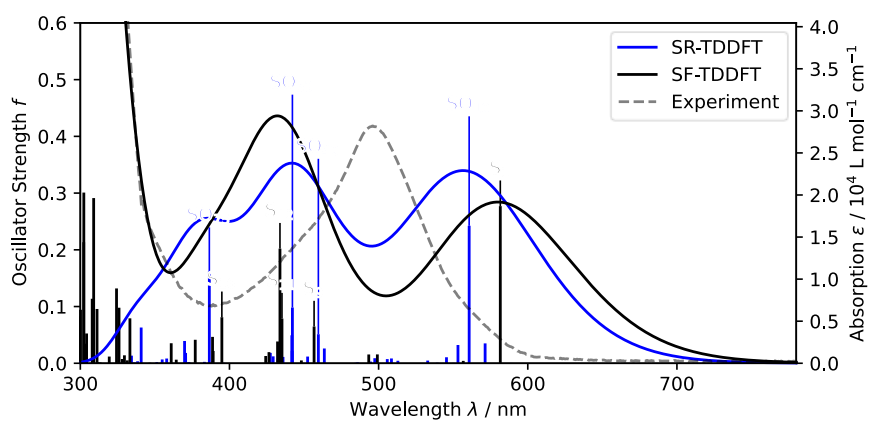

**Figure 2.** Experimental (grey dashed) and simulated UV-vis absorption spectrum of **RuH**. Prominent excitations as obtained by spin-free (SF-)TDDFT (black solid; Gaussian 16) and scalar-relativistic (SR-)TDDFT (blue solid; Orca 5.0) are indicated. Simulated transitions are broadened by Lorentzian functions with a full width at half maximum of 0.2 eV.

In case of the spin-free simulations as performed using Gaussian 16[41], the visible region of the electronic absorption spectrum of **RuH** is dominated by a low-lying and strongly dipole-allowed ILCT excitation at 2.13 eV ($S_1$ at 582 nm) as well as by a set of MLCT transitions. Excitation into the MLCT states $S_9$, $S_{11}$ and $S_{12}$ (at 2.71, 2.85 and 2.86 eV; and 457, 435 and 343 nm, respectively) leads to a population transfer from the $t_{2g}$ orbitals ($d_{xy}$, $d_{xz}$ and $d_{yz}$) into the low-lying $\pi^*_{\text{tpy}}$ orbitals of both coordinating terpyridyl ligands, while $S_{16}$ (3.20 eV; 395 nm) is of mixed ILCT/MLCT character and localizes its excited electron density primarily on the PTZ-tpy ligand. Consistently, the spin-orbit (SO) picture provided by SR-TDDFT reveals one low-lying ILCT transition (into $SO_{10}$) at 2.21 eV (561 nm), which is mainly of $^1$ILCT ($S_1$) character with slight $^3$ILCT, $^3$MLCT and $^3$LLCT contributions associated with $T_3$ and $T_4$ within the spin-free picture, see Table 1.

**Table 1.** Excitation energies (Δ*E*), wavelengths (*λ*) and oscillator strengths (*f*) of prominent singlet-singlet and singlet-triplet excitations contributing to the spin-orbit states in **RuH** (CH$_2$Cl$_2$); simulations performed in Orca 5.0 and Gaussian 16 (in parentheses). Electronic characters are indicated by charge density differences; charge transfer occurs from red to blue.

| singlet-singlet excitations | | | | | singlet-triplet excitations | | | | | spin-orbit excitations | | | | |
|---|---|---|---|---|---|---|---|---|---|---|---|---|---|---|
| state | Δ*E* / eV | *λ* / nm | *f* | character | state | Δ*E* / eV | *λ* / nm | *f* | character | state | composition (weight) | Δ*E* / eV | *λ* / nm | *f* |
| S$_1$ (S$_1$) | 2.20 (2.13) | 563 (582) | 0.315 (0.277) | 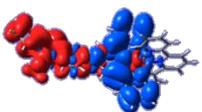 | T$_3$ | 2.18 | 569 | 0.000 | 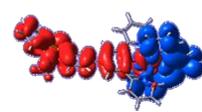 | SO$_{10}$ | S$_1$ (77) T$_4$ (9) T$_3$ (6) | 2.21 | 561 | 0.242 |
| S$_9$ (S$_9$) | 2.65 (2.71) | 467 (457) | 0.058 (0.065) | 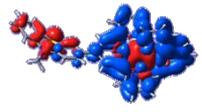 | T$_4$ | 2.23 | 556 | 0.000 | 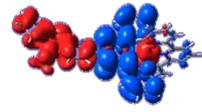 | SO$_{49}$ | T$_{14}$ (40) S$_9$ (28) S$_{10}$ (13) | 2.70 | 460 | 0.051 |
| S$_{10}$ (S$_{11}$) | 2.71 (2.85) | 458 (435) | 0.003 (0.079) | 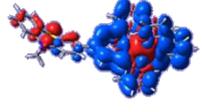 | T$_{14}$ | 2.71 | 457 | 0.000 | 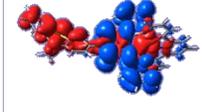 | SO$_{52}$ | S$_{10}$ (64) T$_{14}$ (21) | 2.74 | 452 | 0.012 |
| S$_{11}$ (S$_{12}$) | 2.80 (2.86) | 444 (434) | 0.204 (0.202) | 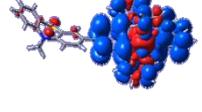 | T$_{15}$ | 2.82 | 440 | 0.000 | 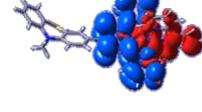 | SO$_{53}$ | S$_{11}$ (49) T$_{15}$ (20) T$_{14}$ (11) T$_{16}$ (7) | 2.80 | 442 | 0.098 |
| S$_{17}$ (S$_{16}$) | 3.20 (3.14) | 388 (395) | 0.166 (0.081) | 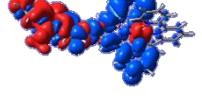 | T$_{16}$ | 2.83 | 438 | 0.000 | 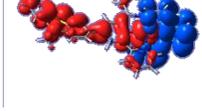 | SO$_{75}$ | S$_{17}$ (98) | 3.21 | 387 | 0.162 |

The higher-lying spin-orbit states $SO_{49}$, and $SO_{52}$ and $SO_{53}$ are of MLCT nature and feature an increased ratio of triplet character. These MLCT transitions can be related to the three MLCT transitions obtained using spin-free TDDFT which involve both tpy ligands ($S_9$, $S_{11}$ and $S_{12}$). Finally, $SO_{75}$ at 3.21 eV (387 nm) is almost entirely of singlet character and represents the formerly discussed mixed ILCT/MLCT transition into $S_{16}$. Thus, spin-free and scalar-relativistic TDDFT simulations draw a consistent picture. Both methods predict a dipole-allowed low-lying ILCT transition at approximately 570 nm as well as several optically active MLCT transitions in the range between 470 and 390 nm. However, the absorption band of the experimental UV-vis spectrum is centred at ~500 nm. Therefore, the TDDFT results are at first glance not in good agreement with the experimental reference. A closer look into the experimental data reveals a weakly absorbing shoulder at approximately 450 nm – in agreement with the MLCT band of the $[Ru(tpy)_2]^{2+}$ parent compound.[86, 87] Therefore, the main absorption band of **RuH** at roughly 500 nm cannot stem exclusively from $MLCT_{tpy}$ transitions. In agreement with the low-lying and optically-allowed ILCT ($\pi_{PTZ}/\pi^*_{tpy}$) and higher-lying $MLCT_{tpy}$ excitations, the resonance Raman data show a decrease in Raman intensity of PTZ-related vibrational modes from 515 to 476 nm and further decrease with 458 nm excitation, while characteristic tpy modes are observed at all three excitation wavelengths.[37] The energetic order of ILCT and MLCT transitions was also confirmed by employing the double-hybrid functional SOS-wPBEPP86,[59] which also predicts the ILCT state well below the MLCT states of interest, see Figure S1a). Unfortunately, all excitation energies are hypsochromically shifted with respect to the experimental data. Finally, the presence of the ILCT state in addition to the MLCT states was confirmed by the restricted active space self-consistent field (RASSCF) method employing a restricted active space (RAS) of (26,2,2;9,6,7), which leads to a multiconfigurational space spanning over almost 700,000 configuration state functions (CSFs), see Figure 3.

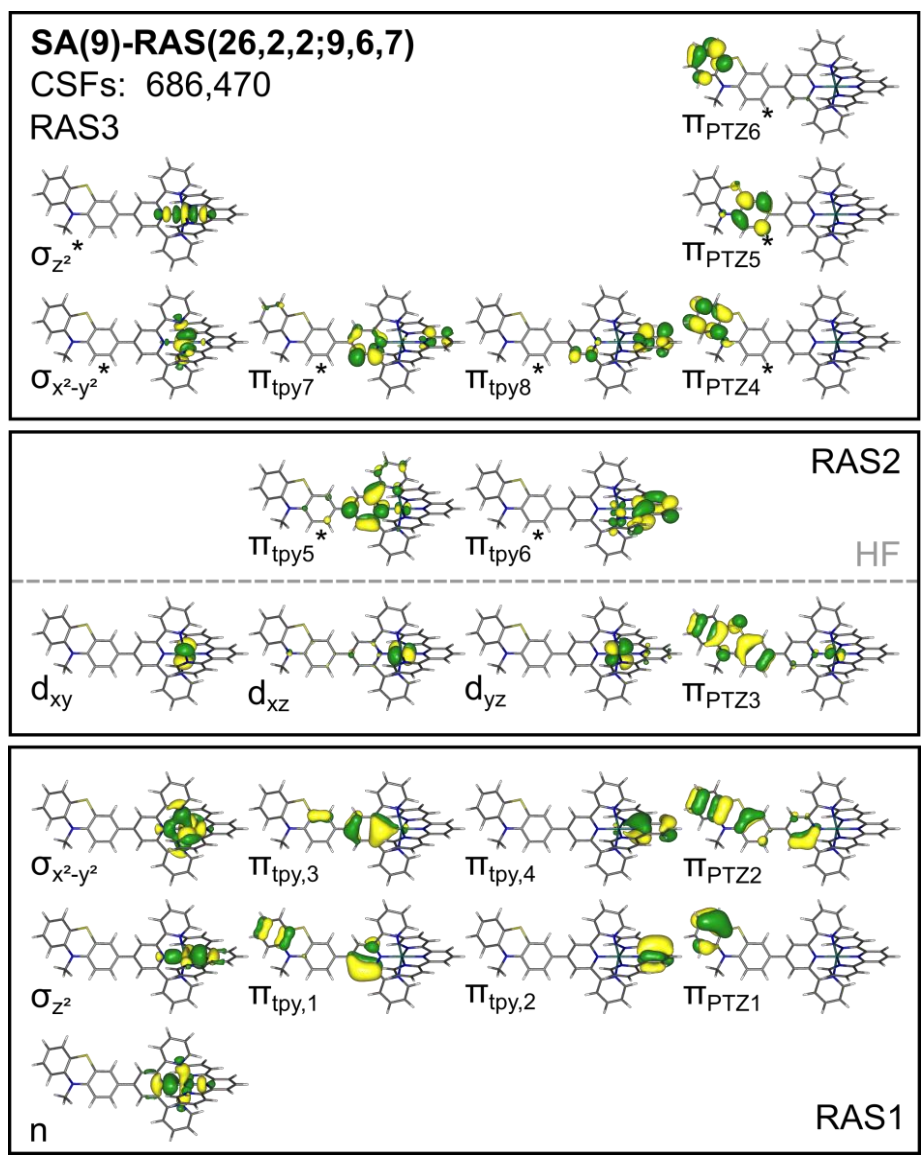

**Figure 3.** Molecular orbitals for the SA(9)-RAS (26,2,2;9,6,7) used in the state average procedure covering the lowest nine singlet roots of **RuH**. The partitioning with respect to the RAS1, RAS2 and RAS3 subspaces as well as the occupation of the molecular orbitals in the Hartree-Fock (HF) reference wavefunction is indicated (grey dashed line). The RAS for the respective triplet state calculations is shown in Figure S3.

The performed state-average procedure takes the nine lowest singlet roots into account and yields two dipole-allowed MLCT excitations ($f$ = 0.905 and 0.084), the one optical accessible ILCT transition of interest ($f$ = 0.146) as well as the singlet ground state and five (mostly) inaccessible states (four MLCTs and one LLCT; $f \approx$ 0). It is noteworthy that RASSCF predicts the ILCT with 4.36 eV at a higher excitation energy than the accessible MLCT states at 2.70, 2.97, 3.01 and 4.05 eV, respectively, (Table S5). However, based on the constructed active space, the MLCT states are better described than the ILCT state as merely one $\pi_{\text{PTZ}}$ orbital is set to the RAS2 subspace. In general, the RASSCF energies are overestimated with respect to the experimental data as well as the TDDFT results. This deviation is not surprising and can be ascribed to the lack of dynamic electron correlation in RASSCF. The dynamic electron correlation can be included upon applying second-order perturbation theory on the RASSCF reference wavefunction (*i.e.*, RASPT2). Unfortunately, multi-state RASPT2 calculations could not be performed for the present transition metal complex due to their enormous computational demand. Furthermore, solvent effects were not included in the multiconfigurational simulations. Detailed information regarding the setup of the multiconfigurational simulations as well as regarding the composition of the active space are presented in the Computational Details (section 2).

In summary, the computational results draw a conclusive picture. The Franck-Condon photophysics of **RuH** are dominated by one low-lying ILCT transition ($\pi_{\text{PTZ}}/\pi^*_{\text{tpy}}$) as well as by several slightly higher-lying MLCT transitions populating $\pi^*_{\text{tpy}}$ orbitals of both terpyridyl ligands.

Finally, the population transfer from the excited singlet states to the triplet manifold upon intersystem crossing (ISC) was exemplarily evaluated for **RuH** by means of SR-TDDFT – employing the scalar relativistic zeroth-order regular approximation (SR-ZORA)[88]. The electronic absorption spectrum as obtained by SR-TDDFT is visualised in Figure 2. As reflected by the composition of the SO transitions discussed above (Table 1), the $^1$ILCT state ($S_1$) is weakly coupled to the triplet states $T_3$ (78 cm$^{-1}$) and $T_4$ (63 cm$^{-1}$) which are of mixed $^3$LLCT/$^3$MLCT and $^3$ILCT/$^3$MLCT character, respectively. Even smaller spin-orbit couplings (SOCs) are predicted with the higher-lying MLCT states $T_{14}$ (12 cm$^{-1}$), $T_{15}$ (18 cm$^{-1}$) and $T_{16}$ (8 cm$^{-1}$). However, significantly stronger SOCs are calculated between the $^1$MLCT states ($S_9$, $S_{10}$, $S_{11}$ and $S_{17}$) and the energetically close-lying triplet states, *i.e.*, triplet states of pronounced $^3$MLCT nature. Of particular interest is the interaction between the strongly dipole-allowed and pure

$^1$MLCT transition (see Orca S$_{11}$ and Gaussian S$_{12}$ in Table 1) and the energetically close triplet states. SR-TDDFT yields SOCs of 216 and 242 cm$^{-1}$ between this $^1$MLCT and the $^3$MLCT states T$_{14}$ and T$_{15}$. Considerably smaller SOCs are observed among the $^1$MLCT states and triplet states of dominant $^3$ILCT or $^3$LLCT (*i.e.*, T$_3$ or T$_4$) character, see Table 2 for details. These findings agree with previous theoretical studies and show that an efficient singlet-to-triplet population transfer is only possible along $^{1/3}$MLCT gateway states.$^{23, 33-35, 89}$

**Table 2.** Spin-orbit coupling elements, $\langle T_j | \hat{H}_{\text{SOC}} | S_i \rangle$ (in cm$^{-1}$), between prominent excited singlet and triplet states of **RuH** in CH$_2$Cl$_2$. All result were obtained by TD-B3LYP as implemented in Orca 5.0.

|  | T$_3$ ($^3$LLCT/$^3$MLCT) | T$_4$ ($^3$ILCT/$^3$MLCT) | T$_{14}$ ($^3$MLCT/$^3$ILCT) | T$_{15}$ ($^3$MLCT/$^3$ILCT) | T$_{16}$ ($^3$MLCT/$^3$LLCT) |
|---|---|---|---|---|---|
| S$_1$ ($^1$ILCT) | 78 | 63 | 12 | 18 | 8 |
| S$_9$ ($^1$MLCT) | 204 | 99 | 88 | 58 | 91 |
| S$_{10}$ ($^1$MLCT) | 57 | 144 | 122 | 171 | 177 |
| S$_{11}$ ($^1$MLCT) | 88 | 84 | 216 | 242 | 152 |
| S$_{17}$ ($^1$ILCT) | 17 | 1 | 14 | 23 | 25 |

Upon thoroughly investigating the Franck-Condon photophysics of the **RuH** parent compound, the TDDFT-based computational protocol was adapted to unravel the nature of the electronic transitions underlying the electronic absorption bands of **RuPh**, **RuTol**, **RuAn** and **RuC$_{60}$** in the visible and UV regions. In agreement with the experimental data, the simulated UV-vis absorption spectra of **RuPh**, **RuTol** and **RuAn** closely resemble the spectrum of the unsubstituted **RuH** species as predicted by TD-B3LYP, see Figure S2a)-c) and Tables S1-3. The $^1$ILCT excitation is consistently predicted to occur in the narrow range between 2.13 and 2.15 eV. This is not surprising as the structural modification is localized at the other terpyridyl ligand. In a similar fashion, the energic positions of these $^1$MLCT states are barely affected by the structural modification. In case of **RuC$_{60}$** the ILCT excitation is split into two transitions (*i.e.*, S$_7$ and S$_8$ at 2.12 and 2.13 eV), due to mixing with an LLCT state from the 10-methylphenothiazinyl moiety to the electron withdrawing fullerene, see Table S4. Likewise, the MLCT properties of **RuC$_{60}$** are slightly altered by the impact of $\pi^*_{\text{C60}}$ orbitals. However, all five complexes feature a low-lying ILCT absorption band and a higher-lying MLCT band in a consistent manner, which was further confirmed by applying the SOS-wPBEPP86 double hybrid functional (Figure S1).

The later discussion will focus on the electron transfer pathways available upon ISC for **RuH**, **RuPh**, **RuTol** and **RuAn**. Due to the influence of the fullerene moiety on the electronic structure of **RuC$_{60}$**, the predicted excited state relaxation pathways and electron transfer kinetics are addressed separately for **RuC$_{60}$**.

### 3.2 Excited-State Electron Transfer Processes

In the following, we investigate the excited state relaxation for the five push-pull triads **RuR** by means of quantum chemical simulations. Herein, we focus on the ET branching channels from the lowest energy $^3$MLCT state (accessible upon ISC, see section 3.1) to the $^3$LLCT as well as to the $^3$ILCT state. A previous experimental investigation based on excited-state spectroscopy and electrochemistry revealed similar thermodynamic properties, *i.e.*, driving forces ($\Delta G$) and reorganization energies ($\lambda$) regarding the electron transfer from a $^3$MLCT donor state to a ligand-based acceptor state for the series of compounds.[37] Experimental evidence also suggests that the introduced substitution pattern allows the electronic coupling ($V_{DA}$) to be remotely controlled for the given pairs of donor and acceptor states in **RuR**. The kinetics of the underlying ET processes are also addressed quantum chemically. Hence, our approach yields driving forces, reorganization energies and electronic couplings based on our lately introduced computational protocol to predict (light-driven) intramolecular charge transfer phenomena in the semi-classical Marcus picture.[38-40] In order to achieve this, the respective $^3$MLCT donor state (D) as well as the ligand-based acceptor states (A; $^3$LLCT and $^3$ILCT) are fully relaxed at the TDDFT level of theory. Subsequently, a linear-interpolated internal coordinate ($R_{ET}$) is constructed to connect $^3$MLCT equilibria with the respective $^3$LLCT and $^3$ILCT structures. Potential energy curves (PECs) are simulated along these efficient coordinates for the states of interest. Subsequently, our lately introduced external optimiser pysisyphus – also aware of excited states – was applied.[74] The diabatization factors were obtained based on the generalized Mulliken-Hush (GMH)[90] method within the crossing region of two diabatic states. Finally, electronic couplings are calculated via the GMH and fragment charge difference (FCD)[91] approaches as well as by virtue of the minimum energy splitting between the involved adiabatic states. Further information on the computational protocol is collected in section 2.

All investigated dyes show only a minor structural rearrangement upon relaxation into the $^3$MLCT donor state equilibrium from the Franck-Condon point. More pronounced structural changes occur upon $^3$LLCT optimization. As shown exemplarily for **RuH** in Figure 4a), structural relaxation mainly involves the PTZ donor moiety. This is attributed to the photooxidation of this group, which leads to a partial planarization in the vicinity of the nitrogen atom, due to the decreased sp$^3$ character. In a similar fashion, structural equilibration of the $^3$ILCT state mainly involves a partial planarization of the PTZ group, see Figure 5a). All optimized and interpolated structures are available from the online repository Zenodo via Ref. [76].

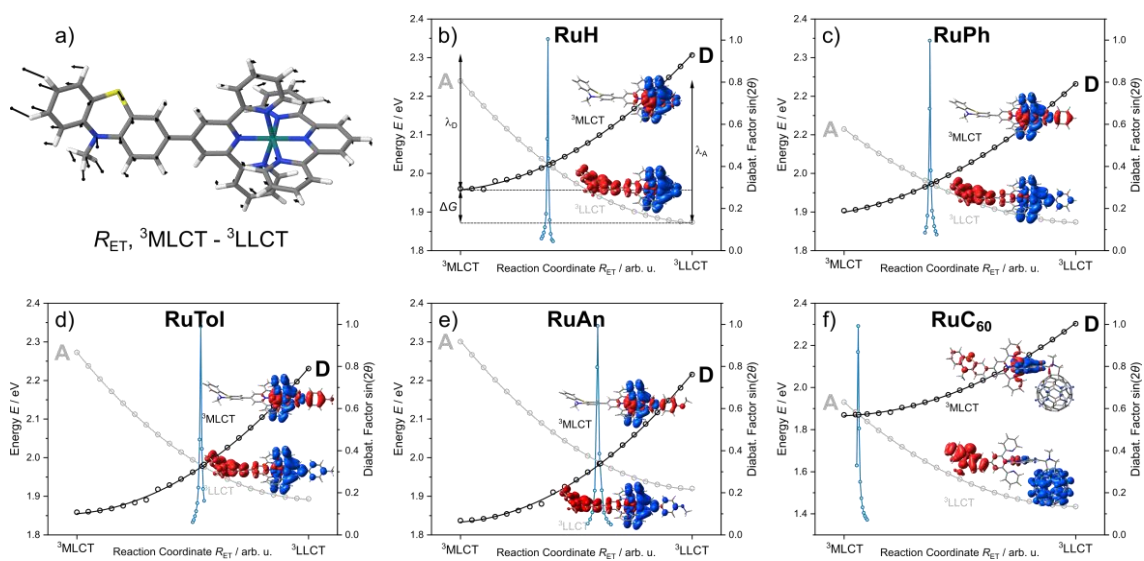

**Figure 4.** a) Linear-interpolated internal coordinate (LIIC, $R_{ET}$) connecting fully relaxed $^3$MLCT and $^3$LLCT structures for **RuH** as shown by displacement vectors. b-f) Calculated diabatic potential energy curves of the $^3$MLCT donor state (D; black) and the $^3$LLCT acceptor state (A; grey) along $R_{ET}$ for **RuH**, **RuPh**, **RuTol**, **RuAn** and **RuC$_{60}$**, respectively. A quadratic polynomial, $E(R_{ET}) = a(R_{ET} - R_0)^2 + E_0$, was fitted to the respective data sets. Diabatization factors (see Equation 4; blue) are illustrated in the crossing region of D and A. Electronic characters for states of interest are visualised by charge density differences (CDDs);

charge transfer occurs from red to blue. Driving force (ΔG) and reorganization energies ($\lambda_D$ and $\lambda_A$) are indicated exemplarily in b).

### 3.2.1 $^3$MLCT to $^3$LLCT channel

Finally, the ET kinetics from the $^3$MLCT donor to the $^3$LLCT acceptor states were evaluated within the semi-classical Marcus picture for the given set of Ru(II)-based compounds. In case of the parent compound **RuH**, a driving force of -0.09 eV is obtained. Very similar reorganization energies of 0.35 and 0.37 eV are predicted for the donor and accept states ($\lambda_D$ and $\lambda_A$), respectively which yields an average reorganization energy ($\lambda_{AVG}$) of 0.36 eV; see Table 3.

**Table 3.** Driving forces (ΔG), reorganization energies (top to bottom: $\lambda_D$, $\lambda_A$ and $\lambda_{AVG}$), electronic couplings ($V_{DA}$), rate constants ($k$) for five **RuR** complexes as obtained at the TDDFT level of theory using the B3LYP hybrid functional and comparison to experimental rate constants.

|  | $^3$MLCT(D) – $^3$LLCT(A) | | | | $^3$MLCT(D) – $^3$ILCT(A) | | | | Exp.[a] |
|---|---|---|---|---|---|---|---|---|---|
|  | ΔG (eV) | $\lambda_i$ (eV) | $V_{DA}$ (eV) | $k_i$ (s$^{-1}$) | ΔG (eV) | $\lambda_i$ (eV) | $V_{DA}$ (eV) | $k_i$ (s$^{-1}$) | $k$ (s$^{-1}$) |
| **RuH** | -0.09 | 0.35 | 4.5x10$^{-4}$ | 7.51x10$^8$ | -0.33 | 0.41 | 1.29x10$^{-2}$ | 3.90x10$^{12}$ | 1.92x10$^{11}$ |
|  |  | 0.37 |  | 6.00x10$^8$ |  | 0.41 |  | 3.87x10$^{12}$ |  |
|  |  | 0.36 |  | 6.71x10$^8$ |  | 0.41 |  | 3.88x10$^{12}$ |  |
| **RuPh** | -0.03 | 0.33 | 4.5x10$^{-4}$ | 3.38x10$^8$ | -0.24 | 0.39 | 1.26x10$^{-2}$ | 2.52x10$^{12}$ | 1.79x10$^{11}$ |
|  |  | 0.24 |  | 9.95x10$^8$ |  | 0.44 |  | 1.68x10$^{12}$ |  |
|  |  | 0.28 |  | 5.77x10$^8$ |  | 0.41 |  | 2.07x10$^{12}$ |  |
| **RuTol** | 0.03 | 0.37 | 7.5x10$^{-4}$ | 1.38x10$^8$ | -0.24 | 0.46 | 1.28x10$^{-2}$ | 1.29x10$^{12}$ | 1.28x10$^{11}$ |
|  |  | 0.38 |  | 1.26x10$^8$ |  | 0.46 |  | 1.32x10$^{12}$ |  |
|  |  | 0.38 |  | 1.31x10$^8$ |  | 0.46 |  | 1.31x10$^{12}$ |  |
| **RuAn** | 0.08 | 0.38 | 11.5x10$^{-4}$ | 8.98x10$^7$ | -0.21 | 0.40 | 1.27x10$^{-2}$ | 1.69x10$^{12}$ | 1.41x10$^{11}$ |
|  |  | 0.38 |  | 8.77x10$^7$ |  | 0.46 |  | 9.70x10$^{11}$ |  |
|  |  | 0.38 |  | 8.87x10$^7$ |  | 0.43 |  | 1.29x10$^{12}$ |  |
| **RuC$_{60}$** | -0.44 | 0.43 | 10.5x10$^{-4}$ | 2.96x10$^{10}$ | -0.30 | 0.43 | 1.05x10$^{-2}$ | 1.97x10$^{12}$ | 4.00x10$^{11}$ |
|  |  | 0.50 |  | 2.55x10$^{10}$ |  | 0.44 |  | 1.82x10$^{12}$ |  |
|  |  | 0.46 |  | 2.80x10$^{10}$ |  | 0.43 |  | 1.90x10$^{12}$ |  |

[a] Experimental electron transfer rates (T = 300 K) taken from Ref. [37].

The almost identical reorganization energies indicate that the parabolic diabatic PECs of **D** and **A** possess curvatures which are a near perfect match. Hence, the PECs are merely displaced along $R_{ET}$ and offset in energy making the applied LIIC a suitable reaction coordinate within the Marcus picture. In addition, similar values of $\lambda_{AVG}$ are obtained for the structurally modified triads **RuPh** (0.28 eV), **RuTol** (0.38 eV) and **RuAn** (0.38 eV). However, a slight systematic decrease of Δ$G$ is predicted with the increase of electron-donating character of the applied substitution pattern at the tpy acceptor ligand. While the $^3$MLCT-$^3$LLCT electron transfer is still slightly exergonic, (Δ$G$ = -0.03 eV) for **RuPh**, the driving force is further decreased to +0.03 and +0.08 eV in case of **RuTol** and **RuAn**, respectively. In general, this range of driving forces ([-0.09 eV; +0.08 eV]) suggests an equilibrium between the two states. In case of **RuC$_{60}$**, the predicted driving force of -0.44 eV is significantly stronger in comparison to the other systems. This is accompanied by an increase of the reorganization energy to roughly 0.46 eV. As mentioned previously, these changes in the thermodynamic properties are related to a pronounced mixing of the $^3$LLCT state (into $\pi^*_{\text{tpy}}$) of interest with a charge transfer to the electron accepting fullerene ($\pi^*_{\text{C60}}$), see Figure 4f).

In addition to Δ$G$ and $\lambda$, the electron transfer processes are governed by the electronic communication between the involved diabatic states described by the electronic coupling ($V_{\text{DA}}$). As shown previously by means of excited state spectroscopy, the applied substitution pattern at the acceptor tpy ligand allows one to remotely control the magnitude of $V_{\text{DA}}$, yielding couplings of 1.17·10$^{-2}$, 1.12·10$^{-2}$, 9.17·10$^{-3}$ and 4.60·10$^{-2}$ eV for **RuH**, **RuPh**, **RuTol** and **RuAn**, respectively. In case of **RuC$_{60}$**, an electronic coupling of 1.95·10$^{-2}$ eV was determined.[37] Computationally, $V_{\text{DA}}$ was obtained for the present set of Ru(II)-based dyes by virtue of the GMH and the FCD approaches, see Equations 3-5 in the Computational Details (section 2). Both methods allow the calculation of $V_{\text{DA}}$ *via* the gap between the adiabatic states and by means of the diabatization factor. The diabatization factor is defined for the states of interest either by the permanent and transition dipoles (GMH) or by the charge distribution (FCD). Figure 4b)-f) depicts the diabatization factors in the vicinity of the crossing region along $R_{\text{ET}}$. Again, the symmetric shape of the calculated diabatization factors suggests that the LIIC is a suitable coordinate to study the present ET reaction. As expected, the largest values (close to 1) are obtained if both states are quasi-degenerate and a mixing angle ($\theta$) of approximately 45° is obtained. Therefore, the electronic coupling at this geometry is governed almost entirely by the energy splitting of the respective adiabatic states (Equation 6). Based on the performed TDDFT simulations and in combination with the GMH methodology, small electronic coupling values of merely $4.5 \cdot 10^{-4}$, $4.5 \cdot 10^{-4}$, $7.5 \cdot 10^{-4}$, $11.5 \cdot 10^{-4}$ and $10.5 \cdot 10^{-4}$ eV are determined for **RuH**, **RuPh**, **RuTol**, **RuAn** and **RuC$_{60}$**, respectively, see Table 3. Numerically identical values were calculated based on the FCD approach (Table S11). The magnitude of $V_{DA}$ indicates a weak interaction between the $^3$MLCT donor and the $^3$LLCT acceptor states, which is attributed to the large distance between the involved redox centers. In agreement with the experimental data, the strongest couplings are predicted for **RuAn** and **RuC$_{60}$**, while **RuH**, **RuPh** and **RuTol** feature rather similar $V_{DA}$ values. However, the theoretically derived electronic couplings are more than one order of magnitude smaller than the experimental values.

Finally, the rate constants for the electron transfer from the $^3$MLCT donor state to the $^3$LLCT acceptor state were calculated within the Marcus picture based on $\Delta G$, $\lambda_{AVG}$ and $V_{DA}$ (Table 3). In case of **RuH** a rate constant of $6.71 \cdot 10^8$ s$^{-1}$ is obtained at the TDDFT level of theory. The magnitude of $k$ indicates a medium-fast ET processes in case of Ru(II)-polypyrdyl-based complexes. For comparison, slightly lower rates of approximately $10^7$ s$^{-1}$ are predicted for $^3$MLCT via triplet metal-centred ($^3$MC) states in Ru(II)-based photocatalysts[40], while much faster ET kinetics of up to ~$10^{13}$ s$^{-1}$ have been reported for $^3$MLCT to $^3$MMCT (metal-to-metal charge transfer) processes in Ru(II)-Co(III) based dyads.[38] While the driving force is slightly decreases from **RuH** to **RuPh**, **RuTol** and **RuAn**, no significant differences are observed for the reorganization energy and the electronic coupling. As a result, the ET rates decrease from **RuH** ($6.71 \cdot 10^8$ s$^{-1}$) to $5.77 \cdot 10^8$, $1.31 \cdot 10^8$, $8.87 \cdot 10^7$ s$^{-1}$ for **RuPh**, **RuTol** and **RuAn**, respectively. In the case of **RuC$_{60}$**, the ET kinetics are faster by two orders of magnitude ($2.80 \cdot 10^{10}$ s$^{-1}$), which is mainly a consequence of the more favorable driving force (recall the influence of the $\pi^*_{C60}$ orbitals). Experimentally, all ET rates were determined to be in the order of magnitude of $10^{11}$ s$^{-1}$ (Table 3). In agreement with the simulated data for the $^3$MLCT-$^3$LLCT channel, the fastest rate was observed for **RuC$_{60}$**. However, in contrast to the computational results, this enhanced rate constant was associated with an increased electronic coupling and not with a larger driving force.

The calculated potential energy curves, electronic couplings and therefore the ET rates are subject to deviations in the TDDFT-predicted excitation energies as obtained by singlet-triplet transitions. Therefore, the PECs and electronic couplings have been evaluated also based on the Tamm-Dancoff approximation (TDA). However, the TDA results are qualitatively identical to the TDDFT results for the present systems, see supporting information and Table 12 for details.

### 3.2.2 $^3$MLCT to $^3$ILCT channel

In an analogous manner to that described above for the $^3$MLCT-$^3$LLCT relaxation pathway, we investigated the second excited-state relaxation channel that is available from the $^3$MLCT donor state, *i.e.*, to a $^3$ILCT acceptor state, see Figure 5.

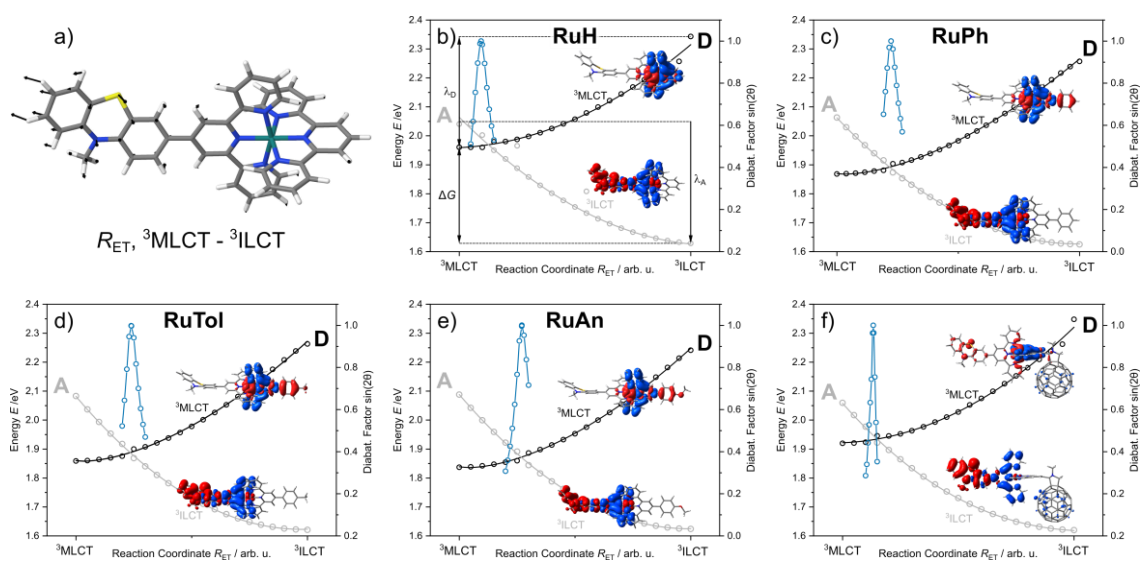

**Figure 5.** a) Linear-interpolated internal coordinate (LIIC, $R_{ET}$) connecting fully relaxed $^3$MLCT and $^3$ILCT structures for **RuH** as shown by displacement vectors. b-f) Calculated diabatic potential energy curves of the $^3$MLCT donor state (D; black) and the $^3$ILCT acceptor state (A; grey) along $R_{ET}$ for **RuH**, **RuPh**, **RuTol**, **RuAn** and **RuC$_{60}$**, respectively. A quadratic polynomial, $E(R_{ET}) = a(R_{ET} - R_0)^2 + E_0$, was fitted to the respective data sets. Diabatization factors (see Equation 4; blue) are illustrated in the crossing region of D and A. Electronic characters for states of interest are visualised by charge density differences (CDDs);

charge transfer occurs from red to blue. Driving force (Δ$G$) and reorganization energies ($\lambda_D$ and $\lambda_A$) are indicated exemplarily in b).

In case of the $^3$MLCT-$^3$ILCT pathway, only minor changes of the driving force are observed for all five Ru(II) complexes, see Table 3. The largest Δ$G$ was obtained for **RuH** (-0.33 eV), decreasing to to -0.24 eV for **RuPh** and **RuTol** and finally to -0.21 eV for **RuAn**. Once again, the more favorable driving force for **RuC$_{60}$** (-0.30 eV) results from the contribution of $\pi^*_{C60}$ orbitals. Likewise, almost identical reorganization energies spanning merely the range from 0.41 to 0.46 eV were predicted at the TDDFT level of theory along the reaction coordinate. The electronic couplings, as obtained by means of the GMH method, were calculated in the crossing region of the diabatic donor ($^3$MLCT) and acceptor ($^3$ILCT) states. The respective diabatization factors are illustrated in Figure 5b)-f). For this pair of redox states, the simulated $V_{DA}$ values are approximately 1.5 orders of magnitude larger than for $^3$MLCT-$^3$LLCT pathway(s) as the involved redox centers are relatively close. In case of **RuH**, a coupling of 1.29·10$^{-2}$ eV was obtained, which is in very good agreement with the experimentally derived $V_{DA}$ of 1.17·10$^{-2}$ eV. However, a pronounced impact of the substitution pattern – as seen by time-resolved experimental spectroscopic techniques – was not observed. In fact, the quantum chemical simulations yield almost identical electronic couplings for **RuPh** (1.29·10$^{-2}$ eV), **RuTol** (1.26·10$^{-2}$ eV) and **RuAn** (1.27·10$^{-2}$ eV). Only in case of the fullerene substituted dye, a slightly smaller value, *i.e.*, 1.05·10$^{-2}$ eV, was calculated. Again, this finding can be explained by the contribution of $\pi^*_{C60}$ orbitals which reduces the overlap of the respective wavefunctions. Due to the favorable driving forces and the larger electronic couplings, the ET kinetics along the $^3$MLCT-$^3$ILCT pathway (~10$^{12}$ s$^{-1}$, Table 3) are by approximately four orders of magnitude faster than the ET kinetics of the previously discussed $^3$MLCT-$^3$LLCT channel. Therefore, the TDDFT-predicted ET rates along the $^3$MLCT-$^3$ILCT cascade are in general in good agreement with the experimental rate constants as obtained by time-resolved spectroscopy (~10$^{11}$ s$^{-1}$, Table 3). However, the marginal variation of the simulated $V_{DA}$ vales along the series of compounds does not allow one to rationalize the trend of the ET rates as observed experimentally.

Finally, the performed quantum chemical simulations map the competitive excited state relaxation cascades as visualized in Figure 6 for **RuH**.

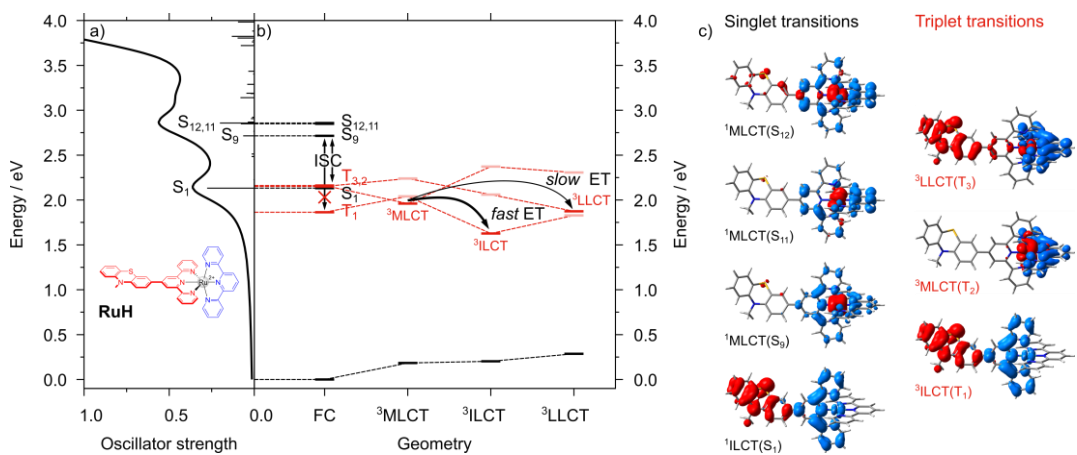

**Figure 6.** a) Simulate UV-vis absorption spectrum of **RuH**. b) Available intersystem crossing (ISC) channels between singlet (in black) and triplet (in red) states of interest within the Franck-Condon (FC) geometry as given by the fully optimized singlet ground state structure ($S_0$). Electron transfer from the lowest equilibrated $^3$MLCT state (donor, $T_2$) to the relaxed $^3$LLCT (acceptor, $T_3$) *vs.* to the fully optimized $^3$ILCT (acceptor, $T_1$) are indicated. The electronic characters of key state involved in the photophysics of **RuH** are shown by charge density differences (CDDs); charge transfer occurs from red to blue.

The initial photoexcitation in the visible spectral region leads to the population of $^1$ILCT ($S_1$) as well as $^1$MLCT ($S_9$, $S_{11}$ and $S_{12}$) excited states. Subsequently, efficient ISC occurs from the $^1$MLCT states to the energetically close-lying $^3$MLCT states, provided by the substantial SOC. Afterwards, internal conversion to the lowest energy $^3$MLCT state follows; see $T_2$ in Figure 6b). Finally, fast ET ($4·10^{12}$ s$^{-1}$) from the equilibrated $^3$MLCT donor state ($T_2$) to the $^3$ILCT acceptor state ($T_1$) is simulated based on semi-classical Marcus theory, while ET along the $^3$MLCT-$^3$LLCT pathway is approximately four orders of magnitude slower ($7·10^8$ s$^{-1}$). In good agreement with the theoretically predicted fast $^3$MLCT-$^3$ILCT conversion, time-resolved absorption studies yielded a slightly slower rate constant of approximately $2·10^{11}$ s$^{-1}$ for the ET from the $^3$MLCT donor state to the ligand-based acceptor state.

Therefore, the performed quantum chemical simulations highlight the importance of the ILCT transition involved in the initial light-activation but also in the (triplet) excited-state relaxation cascade.

## 4 Conclusions

In the present quantum chemical study, a series of five Ru(II)-based photoactive complexes incorporating a 10-methylphenothiazinyl electron donating moiety (**RuH**, **RuPh**, **RuTol**, **RuAn** and **RuC$_{60}$**) were investigated with respect to their excited state properties and relaxation cascades. The strategy of combining inorganic and organic chromophores thereby allows for the localization of the excited electron density away from metal center. As a result, efficient ISC as well as - upon further excited state relaxation - the population of a long-lived triplet state on the organic chromophore are achieved. In particular, the focus was set on addressing the impact of the substitution pattern at the electron accepting terpyridyl ligand. The Franck-Condon photophysics of the parent compound, **RuH**, were thoroughly investigated using time-dependent density functional theory as well as multiconfigurational simulations. These calculations consistently predict a strongly dipole-allowed low-energy ILCT transition to be involved in the light-harvesting, in addition to higher-lying MLCT transitions. However, scalar-relativistic TDDFT clearly reveals that the subsequent intersystem crossing from the accessible singlet states to the triplet manifold proceeds almost exclusively via $^{1/3}$MLCT gateway states. In agreement with previous experimental investigations, the substitution pattern at the accepting terpyridyl ligand has a marginal impact on the Franck-Condon photophysics. However, introduction of the fullerene leads to a pronounced contribution of $\pi^*_{C60}$ orbitals to the electronic transitions.

Subsequently, the electron transfer kinetics of thermally equilibrated lowest-lying $^3$MLCT (donor) state to the $^3$LLCT (acceptor) state *vs.* ET to the $^3$ILCT (acceptor) state were investigated within the semi-classical Marcus picture along a linear-interpolated internal coordinate – connecting the optimized equilibrium structures of the donor and acceptor state. The performed simulations reveal a slow ET from the $^3$MLCT to the

$^3$LLCT state, which is rationalized by the small driving forces in the range of -0.09 to +0.09 eV and the minor electronic coupling between these states (~$10^{-4}$ eV). Contrastingly, the favourable $\Delta G$ ([-0.3 eV: -0.2 eV]) and $V_{DA}$ values (~$10^{-2}$ eV) for $^3$MLCT-$^3$ILCT process lead to fast ET processes along this pathway. Thus, the theoretical results clearly reveal that charge separation occurs on the PTZ-substituted terpyridine ligand. Only in case of **RuC$_{60}$**, competitive charge-separation processes might take place as population of the $^3$ILCT state is only two orders of magnitude slower than the population of the $^3$LLCT state.

On one hand, the performed quantum chemical simulations could rationalize – in agreement with previous experimental studies – a pronounced effect and trend of the substitution pattern on the underlying electronic coupling along the $^3$MLCT-$^3$LLCT channel, the absolute value of these couplings deviate by more than one order of magnitude. On the other hand, the absolute values of the coupling along the $^3$MLCT-$^3$ILCT pathway were determined to be independent with respect to structural modifications, yet in magnitude in excellent agreement with the experimental observations, *i.e.*, for **RuH** 1.29·$10^{-2}$ eV (theory) *vs.* 1.17·$10^{-2}$ eV (experiment). We speculate that the measured remote control effect of the structural substitution pattern on the electronic coupling might originate from interference effects between the involved excited states, in particular between the $^3$MLCT, the $^3$LLCT and the $^3$ILCT states.

Future computational studies will elaborate on different methods to approximate the reaction coordinate in the excited state. Further, we will evaluate such possible interference effects between these three states of interest ($^3$MLCT, $^3$LLCT and $^3$ILCT) using dissipative quantum dynamics simulations that also allow the description of superexchange phenomena and incomplete population transfer. Furthermore, the capability to tune the electron transfer processes by virtue of the underlying electronic coupling will be explored for structurally closely related systems in a joint synthetic-spectroscopic-theoretical fashion.

**Supplementary material**

See the supplementary material for details on spin-free and spin-orbit states contributing to the Franck-Condon photophysics of all complexes, multiconfigurational simulations, and electron transfer kinetics as obtain using the Tamm-Dancoff approximation.

**Acknowledgements**

We thank Philipp Traber for scientific discussions. Financial support by the Deutsche Forschungsgemeinschaft (DFG, German Research Foundation) – Projektnummers 456209398 and 448713509 as well as Projektnummer 364549901, TRR 234 [A1 and A4] – is gratefully acknowledged. All calculations were performed at the Universitätsrechenzentrum of the Friedrich-Schiller-University Jena.

**Conflict of interest**

The authors have no conflicts to disclose.

**Author contributions**

S.K. and B.D.-I. performed the conceptualization and the project-administration. G.Y. and C.Z. performed quantum chemical simulations. G.Y. and S.K. prepared the figures, the writing of the original draft was performed by G.Y., G.E.S. and S.K. All authors contributed to the writing by reviewing and editing of the original draft.

**Guangjun Yang**: Conceptualization (equal); Data curation (lead); Formal analysis (lead); Visualization (lead); Writing – original draft (lead); Writing – review & editing (lead). **Georgina E. Shillito**: Conceptualization (support); Supervision (support); Formal analysis (equal); Visualization (support); Writing – original draft (lead); Writing – review & editing (equal). **Clara Zens**: Conceptualization (support); Data curation (equal); Formal analysis (equal); Visualization (support); Writing – original draft (support); Writing – review & editing (equal). **Benjamin Dietzek-Ivanšić**: Conceptualization (lead); Funding acquisition (lead); Project administration (lead); Resources (lead); Supervision (support); Visualization (support); Writing – original draft (support); Writing – review & editing (lead). **Stephan Kupfer**: Conceptualization (lead); Funding acquisition (lead); Project administration (lead); Resources (lead); Supervision (lead); Visualization (equal); Writing – original draft (lead); Writing – review & editing (lead).

**Data availability**

The data that support the findings of this study are available from the corresponding author upon reasonable request. All structures are available from the online repository Zenodo via Ref. [76].

**Entry for the Table of Contents**

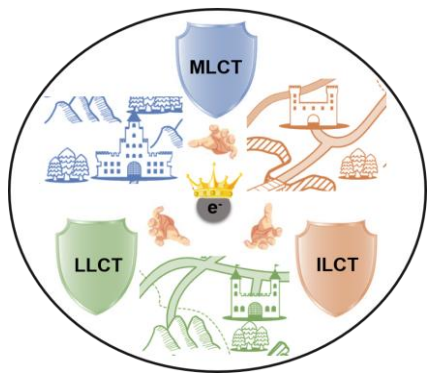

**Supporting Information**

**The three kingdoms – Photoinduced electron transfer cascades controlled by electronic couplings**

Guangjun Yang,[a] Georgina E. Shillito,[a] Clara Zens,[a] Benjamin Dietzek-Ivanšić,[a,b] Stephan Kupfer[a]*

Dedicated to Prof. Wolfgang Weigand on occasion of his 65[th] birthday.

[a] G. Yang, C. Zens, Dr. G. E. Shillito, Dr. S. Kupfer, Prof. Dr. B. Dietzek-Ivanšić

Institute of Physical Chemistry, Friedrich Schiller University Jena,

Helmholtzweg 4, 07743 Jena, Germany

E-mail: stephan.kupfer@uni-jena.de

[b] Prof. Dr. B. Dietzek-Ivanšić

Leibniz Institute of Photonic Technology (IPHT) e.V. Department Functional Interfaces,

Albert-Einstein-Straße 9, 07745 Jena, Germany

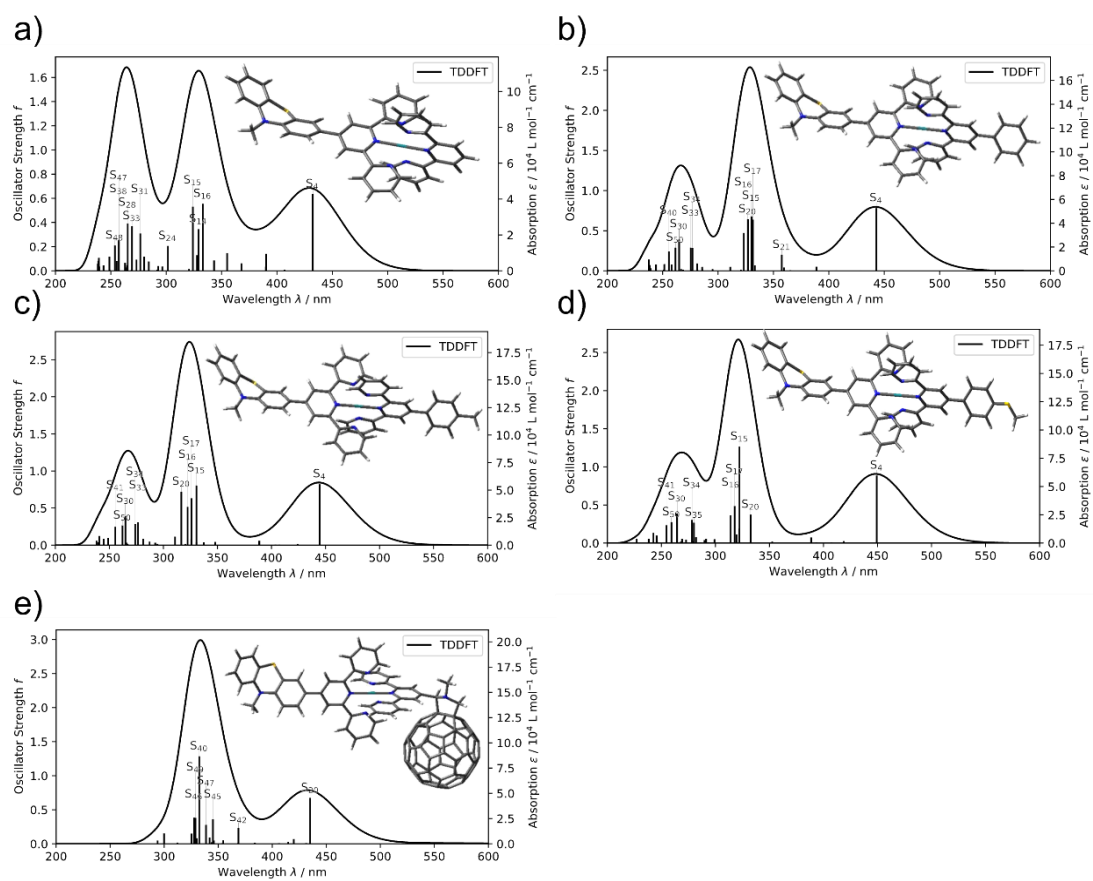

**Figure S1.** Simulated absorption spectra of **RuR** complexes (a-e) obtained by means of TDDFT using the double-hybrid functional SOS-wPBEPP86 as implemented in Orca. Simulated transitions are broadened by Lorentzian functions with a full width at half maximum of 0.2 eV.

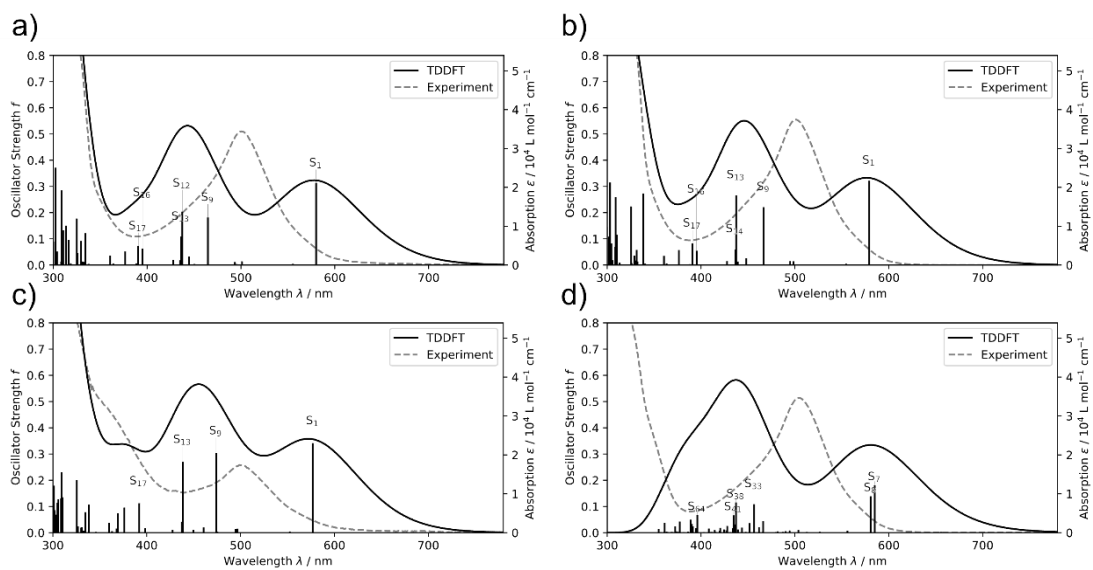

**Figure S2.** Experimental absorption spectrum (dashed grey), simulated absorption spectra of **RuPh**, **RuTol**, **RuAn** and **RuC$_{60}$**, (a-d) obtained by means of TD-B3LYP as implemented in Gaussian 16 (solid black). Simulated transitions are broadened by Lorentzian functions with a full width at half maximum of 0.2 eV.

**Table S1.** Simulated excited state properties of the low-lying bright singlet excited states of **RuPh** in $CH_2Cl_2$ such as excitation energies (in eV), excitation wave lengths (in nm), oscillator strengths, spin contamination, MO pairs, leading transitions as represented by charge density differences (CDDs; charge transfer takes place from red to blue). All results were obtained using the B3LYP functional as implemented in Gaussian 16.

| Transition ($S_0 \rightarrow S_x$) | $\Delta E$ / eV | $\lambda$ / nm | $f$ | $\langle s^2 \rangle$ | Character |
|---|---|---|---|---|---|
| $S_1$ | 2.14 | 580 | 0.3135 | 0.000 | 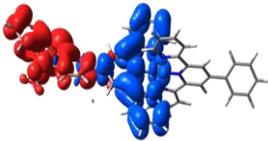 |
| $S_9$ | 2.67 | 465 | 0.1817 | 0.000 | 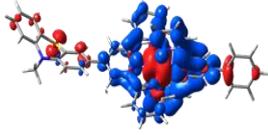 |
| $S_{12}$ | 2.83 | 437 | 0.2038 | 0.000 | 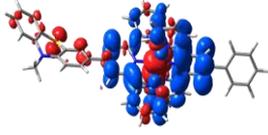 |
| $S_{13}$ | 2.84 | 436 | 0.1090 | 0.000 | 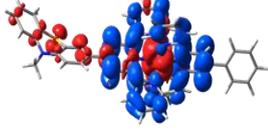 |
| $S_{16}$ | 3.14 | 395 | 0.0628 | 0.000 | 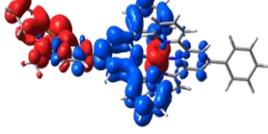 |
| $S_{17}$ | 3.17 | 391 | 0.0730 | 0.000 | 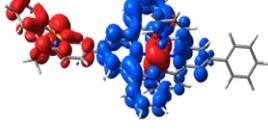 |

**Table S2.** Simulated excited state properties of the low-lying bright singlet excited states of **RuTol** in $CH_2Cl_2$ such as excitation energies (in eV), excitation wave lengths (in nm), oscillator strengths, spin contamination, MO pairs, leading transitions as represented by charge density differences (CDDs; charge transfer takes place from red to blue). All results were obtained using the B3LYP functional as implemented in Gaussian 16.

| Transition ($S_0 \rightarrow S_x$) | $\Delta E$ / eV | $\lambda$ / nm | $f$ | $\langle s^2 \rangle$ | Character |
|---|---|---|---|---|---|
| $S_1$ | 2.14 | 579 | 0.3218 | 0.000 | 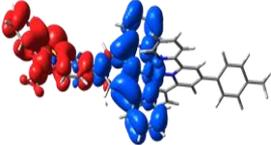 |
| $S_9$ | 2.66 | 467 | 0.2211 | 0.000 | 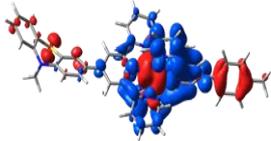 |
| $S_{13}$ | 2.83 | 437 | 0.2664 | 0.000 | 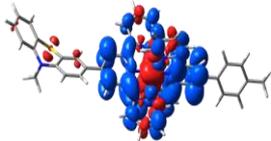 |
| $S_{14}$ | 2.84 | 436 | 0.0601 | 0.000 | 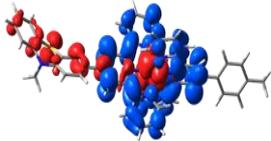 |
| $S_{16}$ | 3.14 | 395 | 0.0544 | 0.000 | 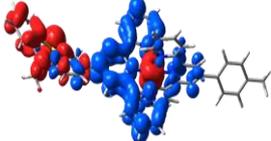 |
| $S_{17}$ | 3.17 | 391 | 0.0820 | 0.000 | 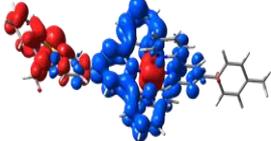 |

**Table S3.** Simulated excited state properties of the low-lying bright singlet excited states of **RuAn** in $CH_2Cl_2$ such as excitation energies (in eV), excitation wave lengths (in nm), oscillator strengths, spin contamination, MO pairs, leading transitions as represented by charge density differences (CDDs; charge transfer takes place from red to blue). All results were obtained using the B3LYP functional as implemented in Gaussian 16.

| Transition ($S_0 \rightarrow S_x$) | $\Delta E$ / eV | $\lambda$ / nm | $f$ | $\langle s^2 \rangle$ | Character |
|---|---|---|---|---|---|
| $S_1$ | 2.15 | 577 | 0.3415 | 0.000 | 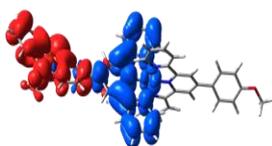 |
| $S_9$ | 2.62 | 474 | 0.3033 | 0.000 | 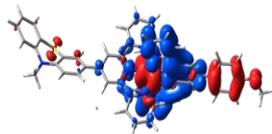 |
| $S_{13}$ | 2.83 | 438 | 0.2702 | 0.000 | 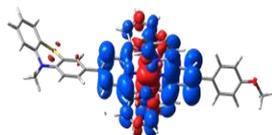 |
| $S_{17}$ | 3.16 | 392 | 0.1122 | 0.000 | 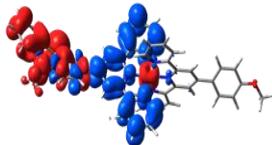 |

**Table S4.** Simulated excited state properties of the low-lying bright singlet excited states of **RuC$_{60}$** in CH$_2$Cl$_2$ such as excitation energies (in eV), excitation wave lengths (in nm), oscillator strengths, spin contamination, MO pairs, leading transitions as represented by charge density differences (CDDs; charge transfer takes place from red to blue). All results were obtained using the B3LYP functional as implemented in Gaussian 16.

| Transition (S$_0$ → S$_x$) | ΔE / eV | λ / nm | f | ⟨s²⟩ | Character |
|---|---|---|---|---|---|
| S$_7$ | 2.12 | 585 | 0.1807 | 0.000 | 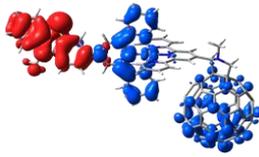 |
| S$_8$ | 2.13 | 581 | 0.1389 | 0.000 | 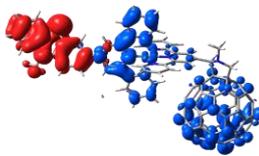 |
| S$_{33}$ | 2.72 | 456 | 0.1087 | 0.000 | 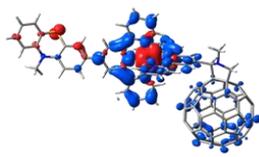 |
| S$_{38}$ | 2.84 | 437 | 0.1153 | 0.000 | 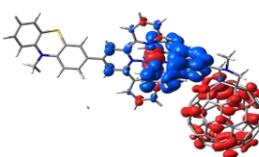 |
| S$_{41}$ | 2.85 | 435 | 0.0664 | 0.000 | 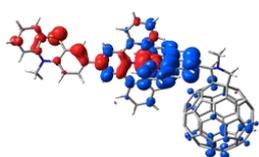 |
| S$_{64}$ | 3.13 | 396 | 0.0672 | 0.000 | 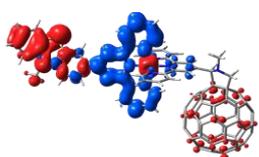 |

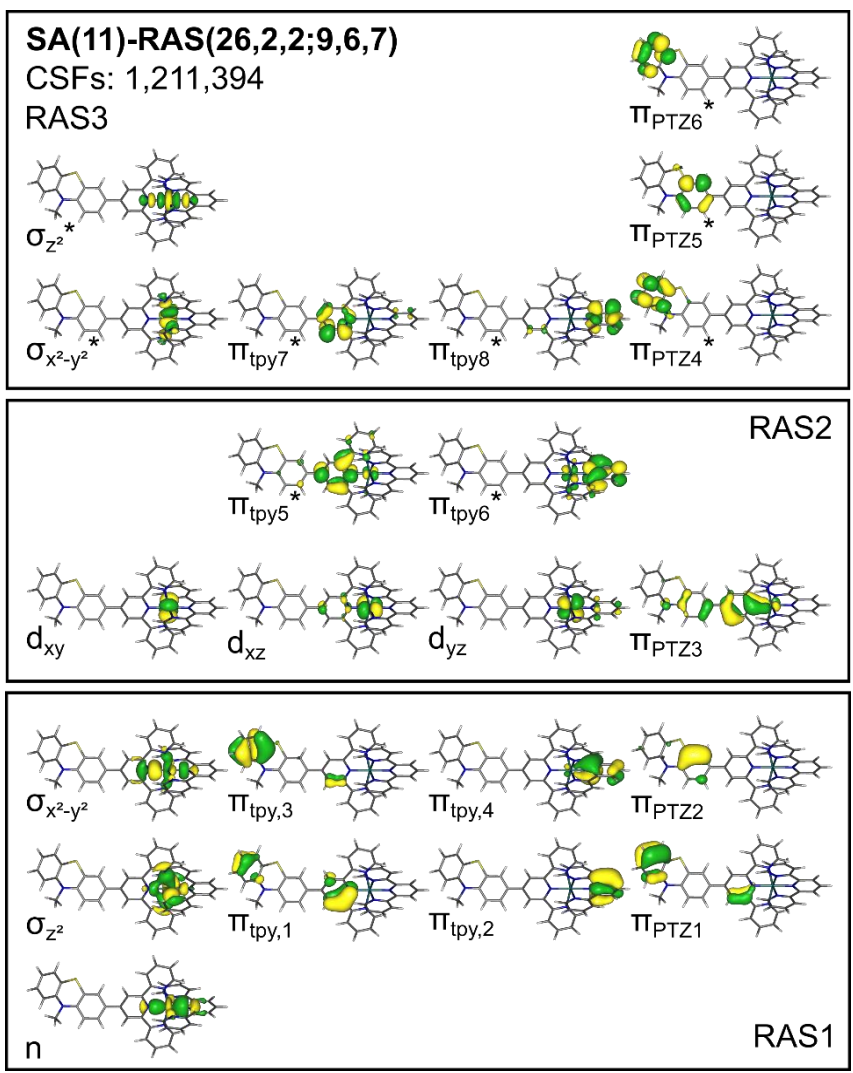

**Figure S3.** Molecular orbitals for the SA(11)-RAS (26,2,2;9,6,7) used in the state average procedure covering the lowest eleven triplet roots of **RuH**. The partitioning with respect to the RAS1, RAS2 and RAS3 subspaces as well as the occupation of the molecular orbitals in the Hartree-Fock (HF) reference wavefunction is indicated (grey dashed line). The RAS for the respective singlet state calculations is shown in Figure 3.

**Table S5.** Simulated gas phase excited state properties of the low-lying singlet (state-average 9) and triplet (state-average 11) roots of **RuH** such as excitation energies (in eV), excitation wave lengths (in nm) and oscillator strengths. Leading configurations (≥ 5%) are given with respect to the closed-shell Hartree-Fock (HF) reference; double excitations are labeled DE. Excitation energies and oscillator strengths in $CH_2Cl_2$ are approximated based on the respective TDDFT data and indicated in parenthesis.

| | Singlet: SA(9)-RAS(26,2,2;9,6,7) | | | | | |
|---|---|---|---|---|---|---|
| | CSFs: 686,470 | | | | | |
| Root | Configuration | Weight / % | Character | $\Delta E$ / eV | $\lambda$ / nm | $f$ |
| 1 | HF | 77 | - | 0 | - | - |
| | $d_{yz} \rightarrow \pi^*_{tpy6}$ | 5 | MLCT$_{tpy}$ | | | |
| 2 | $d_{xy} \rightarrow \pi^*_{tpy6}$ | 82 | MLCT$_{tpy}$ | 2.70 (2.75) | 459 | 0.0043 (0.0044) |
| 3 | $d_{xz} \rightarrow \pi^*_{tpy6}$ | 80 | MLCT$_{tpy}$ | 2.77 (2.67) | 447 | 0.0000 (0.0000) |
| 4 | $d_{yz} \rightarrow \pi^*_{tpy6}$ | 46 | MLCT$_{tpy}$ | 2.97 (2.85) | 418 | 0.9046 (0.8682) |
| | $d_{xz} \rightarrow \pi^*_{tpy5}$ | 36 | MLCT$_{tpy}$ | | | |
| 5 | $d_{xy} \rightarrow \pi^*_{tpy5}$ | 81 | MLCT$_{tpy}$ | 3.01 (2.89) | 412 | 0.0021 (0.0021) |
| 6 | $d_{yz} \rightarrow \pi^*_{tpy5}$ | 84 | MLCT$_{tpy}$ | 3.18 (3.08) | 390 | 0.0000 (0.0000) |
| 7 | $d_{xz} \rightarrow \pi^*_{tpy5}$ | 42 | MLCT$_{tpy}$ | 4.05 (3.95) | 306 | 0.0484 (0.0472) |
| | $d_{yz} \rightarrow \pi^*_{tpy6}$ | 31 | MLCT$_{tpy}$ | | | |
| | HF | 7 | HF | | | |
| 8 | $\pi_{PTZ3} \rightarrow \pi^*_{tpy5}$ | 62 | ILCT | 4.36 (5.30) | 284 | 0.1458 (0.1772) |
| | $d_{yz}, \pi_{PTZ3} \rightarrow \pi^*_{tpy5}, \pi^*_{tpy6}$ | 11 | DE | | | |
| 9 | $\pi_{PTZ3} \rightarrow \pi^*_{tpy6}$ | 48 | LLCT | 4.71 (5.88) | 263 | 0.0000 (0.0000) |
| | $d_{xz}, \pi_{PTZ3} \rightarrow \pi^*_{tpy5}, \pi^*_{tpy6}$ | 30 | DE | | | |
| | Triplet: SA(11)-RAS(26,2,2;9,6,7) | | | | | |
| | CSFs: 1,211,394 | | | | | |
| Root | Configuration | Weight / % | Character | $\Delta E$ / eV | $\lambda$ / nm | $f$ |
| 1 | $d_{yz} \rightarrow \pi^*_{tpy6}$ | 81 | MLCT$_{tpy}$ | 2.30 (2.36) | 539 | - |
| 2 | $d_{xz} \rightarrow \pi^*_{tpy5}$ | 77 | MLCT$_{tpy}$ | 2.52 (2.52) | 493 | - |
| 3 | $d_{xy} \rightarrow \pi^*_{tpy6}$ | 79 | MLCT$_{tpy}$ | 2.54 (2.60) | 489 | - |
| 4 | $d_{xz} \rightarrow \pi^*_{tpy6}$ | 79 | MLCT$_{tpy}$ | 2.55 (2.62) | 486 | - |
| 5 | $d_{xy} \rightarrow \pi^*_{tpy5}$ | 81 | MLCT$_{tpy}$ | 3.02 (2.91) | 410 | - |
| 6 | $d_{yz} \rightarrow \pi^*_{tpy5}$ | 82 | MLCT$_{tpy}$ | 3.17 (3.01) | 392 | - |
| 7 | $d_{xy} \rightarrow \sigma^*_{x^2-y^2}$ | 84 | MC | 4.35 (4.36) | 285 | - |
| 8 | $\pi_{PTZ3} \rightarrow \pi^*_{tpy5}$ | 68 | ILCT | 4.40 (5.33) | 282 | - |
| 9 | $d_{xz} \rightarrow \sigma^*_{x^2-y^2}$ | 80 | MC | 5.09 (5.24) | 244 | - |
| 10 | $d_{yz} \rightarrow \sigma^*_{x^2-y^2}$ | 85 | MC | 5.22 (5.22) | 238 | - |
| 11 | $\pi_{PTZ3} \rightarrow \pi^*_{tpy6}$ | 33 | LLCT | 5.58 (6.72) | 222 | - |
| | $d_{xz}, \pi_{PTZ3} \rightarrow \pi^*_{tpy5}, \pi^*_{tpy6}$ | 28 | DE | | | |
| | $d_{xz}, d_{xz} \rightarrow \pi^*_{tpy5}, \pi^*_{tpy6}$ | 16 | DE | | | |

**Table S6.** Simulated excited state properties of the low-lying bright singlet excited states of **RuH** in CH$_2$Cl$_2$ such as excitation energies (in eV), excitation wave lengths (in nm), oscillator strengths, spin contamination, MO pairs, leading transitions as represented by charge density differences (CDDs; charge transfer takes place from red to blue). All results were obtained using the B3LYP functional as implemented in Orca 5.0.

| Transition (S$_0$ → S$_x$) | Δ$E$ / eV | $\lambda$ / nm | $f$ | $\langle s^2 \rangle$ | Character |
|---|---|---|---|---|---|
| S$_4$ | 2.87 | 432 | 0.6355 | 0.000 | 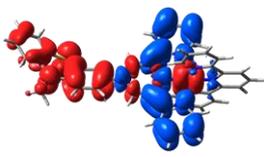 |
| S$_{14}$ | 3.77 | 329 | 0.3450 | 0.000 | 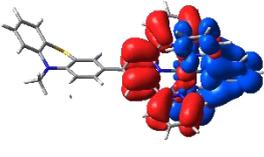 |
| S$_{15}$ | 3.83 | 324 | 0.5307 | 0.000 | 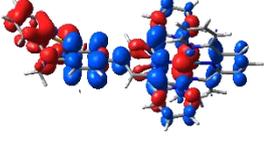 |
| S$_{16}$ | 3.72 | 333 | 0.5549 | 0.000 | 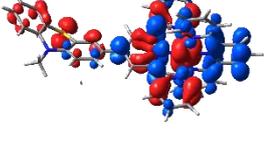 |
| S$_{24}$ | 4.11 | 302 | 0.2057 | 0.000 | 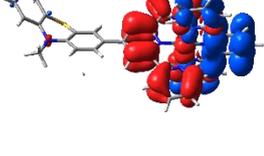 |
| S$_{28}$ | 4.67 | 265 | 0.3913 | 0.000 | 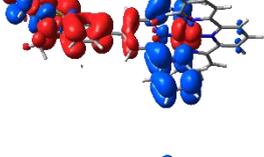 |
| S$_{31}$ | 4.48 | 277 | 0.3084 | 0.000 | 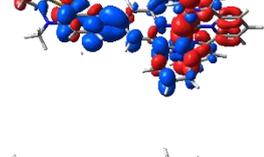 |
| S$_{33}$ | 4.61 | 269 | 0.3704 | 0.000 | 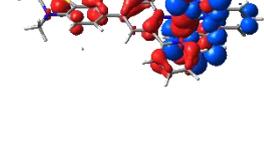 |

| Transition (S₀ → Sₓ) | ΔE / eV | λ / nm | f | ⟨s²⟩ | Character |
|---|---|---|---|---|---|
| S₃₈ | 4.82 | 257 | 0.2535 | 0.000 | 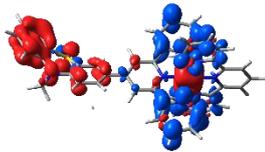 |
| S₄₇ | 4.82 | 257 | 0.2412 | 0.000 | 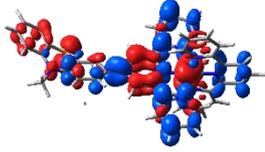 |
| S₄₈ | 4.89 | 254 | 0.2098 | 0.000 | 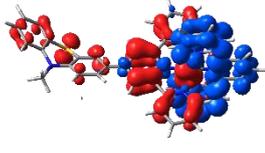 |

**Table S7.** Simulated excited state properties of the low-lying bright singlet excited states of **RuPh** in CH₂Cl₂ such as excitation energies (in eV), excitation wave lengths (in nm), oscillator strengths, spin contamination, MO pairs, leading transitions as represented by charge density differences (CDDs; charge transfer takes place from red to blue). All results were obtained using the B3LYP functional as implemented in Orca 5.0.

| Transition (S₀ → Sₓ) | ΔE / eV | λ / nm | f | ⟨s²⟩ | Character |
|---|---|---|---|---|---|
| S₄ | 2.80 | 443 | 0.7824 | 0.000 | 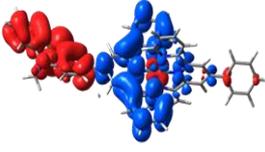 |
| S₁₅ | 3.76 | 330 | 0.6785 | 0.000 | 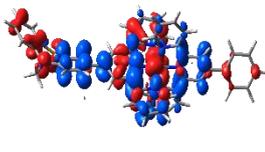 |
| S₁₆ | 3.84 | 323 | 0.4712 | 0.000 | 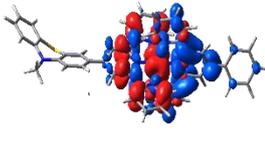 |
| S₁₇ | 3.74 | 331 | 0.6391 | 0.000 | 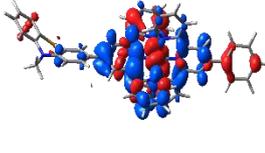 |

| Transition (S$_0$ → S$_x$) | ΔE / eV | λ / nm | f | ⟨s²⟩ | Character |
|---|---|---|---|---|---|
| S$_{20}$ | 3.79 | 327 | 0.6424 | 0.000 | 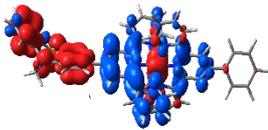 |
| S$_{21}$ | 3.47 | 358 | 0.2002 | 0.000 | 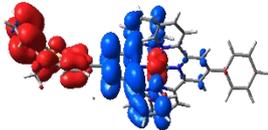 |
| S$_{30}$ | 4.68 | 265 | 0.3886 | 0.000 | 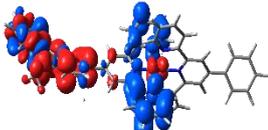 |
| S$_{33}$ | 4.50 | 275 | 0.2894 | 0.000 | 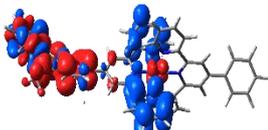 |
| S$_{34}$ | 4.48 | 277 | 0.2839 | 0.000 | 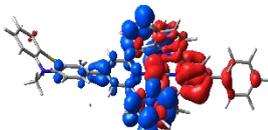 |
| S$_{40}$ | 4.85 | 256 | 0.2433 | 0.000 | 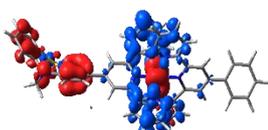 |
| S$_{50}$ | 4.74 | 262 | 0.2895 | 0.000 | 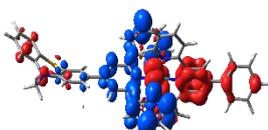 |

**Table S8.** Simulated excited state properties of the low-lying bright singlet excited states of **RuTol** in $CH_2Cl_2$ such as excitation energies (in eV), excitation wave lengths (in nm), oscillator strengths, spin contamination, MO pairs, leading transitions as represented by charge density differences (CDDs; charge transfer takes place from red to blue). All results were obtained using the B3LYP functional as implemented in Orca 5.0.

| Transition (S$_0$ → S$_x$) | ΔE / eV | λ / nm | f | ⟨s²⟩ | Character |
|---|---|---|---|---|---|
| S$_4$ | 2.79 | 445 | 0.8225 | 0.000 | 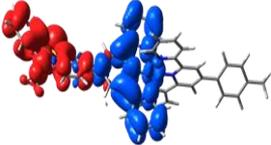 |
| S$_{15}$ | 3.75 | 331 | 0.8032 | 0.000 | 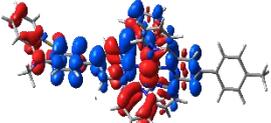 |
| S$_{16}$ | 3.85 | 332 | 0.5149 | 0.000 | 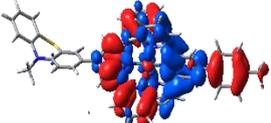 |
| S$_{17}$ | 3.80 | 326 | 0.6313 | 0.000 | 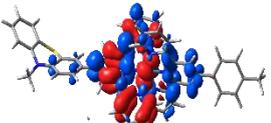 |
| S$_{20}$ | 3.92 | 317 | 0.7211 | 0.000 | 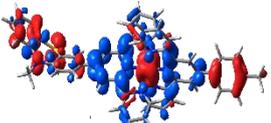 |
| S$_{30}$ | 4.68 | 265 | 0.3900 | 0.000 | 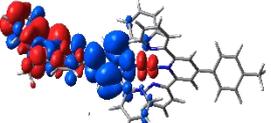 |
| S$_{33}$ | 4.49 | 277 | 0.3076 | 0.000 | 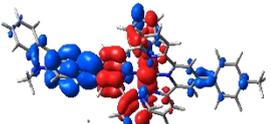 |
| S$_{34}$ | 4.53 | 274 | 0.2862 | 0.000 | 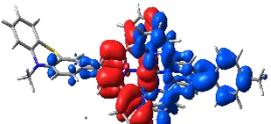 |

| Transition ($S_0 \rightarrow S_x$) | $\Delta E$ / eV | $\lambda$ / nm | $f$ | $\langle s^2 \rangle$ | Character |
|---|---|---|---|---|---|
| $S_{41}$ | 4.85 | 256 | 0.2466 | 0.000 | 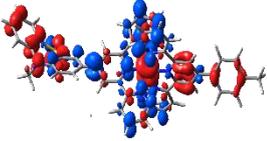 |
| $S_{50}$ | 4.73 | 262 | 0.2655 | 0.000 | 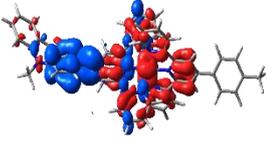 |

**Table S9.** Simulated excited state properties of the low-lying bright singlet excited states of **RuAn** in $CH_2Cl_2$ such as excitation energies (in eV), excitation wave lengths (in nm), oscillator strengths, spin contamination, MO pairs, leading transitions as represented by charge density differences (CDDs; charge transfer takes place from red to blue). All results were obtained using the B3LYP functional as implemented in Orca 5.0.

| Transition ($S_0 \rightarrow S_x$) | $\Delta E$ / eV | $\lambda$ / nm | $f$ | $\langle s^2 \rangle$ | Character |
|---|---|---|---|---|---|
| $S_4$ | 2.76 | 449 | 0.8863 | 0.000 | 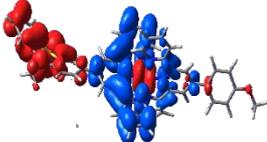 |
| $S_{15}$ | 3.85 | 322 | 1.2662 | 0.000 | 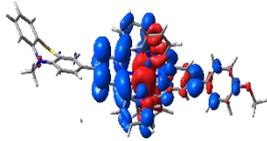 |
| $S_{16}$ | 3.95 | 314 | 0.3646 | 0.000 | 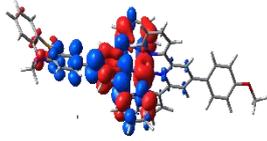 |
| $S_{17}$ | 3.90 | 318 | 0.4851 | 0.000 | 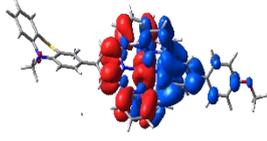 |
| $S_{20}$ | 3.73 | 333 | 0.3737 | 0.000 | 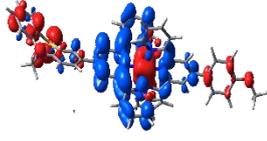 |

| Transition ($S_0 \to S_x$) | $\Delta E$ / eV | $\lambda$ / nm | $f$ | $\langle s^2 \rangle$ | Character |
|---|---|---|---|---|---|
| $S_{30}$ | 4.69 | 265 | 0.3888 | 0.000 | 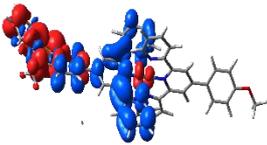 |
| $S_{34}$ | 4.46 | 278 | 0.3045 | 0.000 | 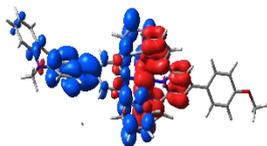 |
| $S_{35}$ | 4.43 | 280 | 0.2682 | 0.000 | 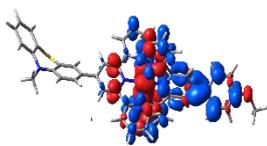 |
| $S_{41}$ | 4.87 | 255 | 0.2339 | 0.000 | 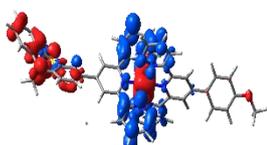 |
| $S_{50}$ | 4.78 | 259 | 0.2756 | 0.000 | 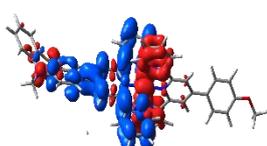 |

**Table S10.** Simulated excited state properties of the low-lying bright singlet excited states of **RuC$_{60}$** in CH$_2$Cl$_2$ such as excitation energies (in eV), excitation wave lengths (in nm), oscillator strengths, spin contamination, MO pairs, leading transitions as represented by charge density differences (CDDs; charge transfer takes place from red to blue). All results were obtained using the B3LYP functional as implemented in Orca 5.0.

| Transition ($S_0 \to S_x$) | $\Delta E$ / eV | $\lambda$ / nm | $f$ | $\langle s^2 \rangle$ | Character |
|---|---|---|---|---|---|
| $S_{20}$ | 2.85 | 435 | 0.6742 | 0.000 | 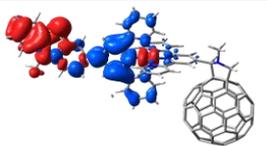 |
| $S_{40}$ | 3.73 | 332 | 1.2867 | 0.000 | 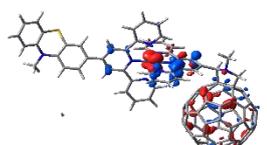 |

| Transition (S₀ → Sₓ) | $\Delta E$ / eV | $\lambda$ / nm | $f$ | $\langle s^2 \rangle$ | Character |
|---|---|---|---|---|---|
| $S_{42}$ | 3.36 | 369 | 0.2332 | 0.000 | 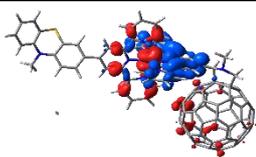 |
| $S_{45}$ | 3.59 | 345 | 0.3604 | 0.000 | 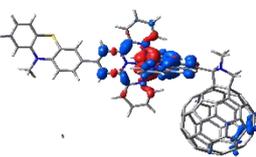 |
| $S_{46}$ | 3.79 | 328 | 0.3838 | 0.000 | 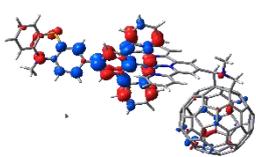 |
| $S_{47}$ | 3.66 | 339 | 0.2778 | 0.000 | 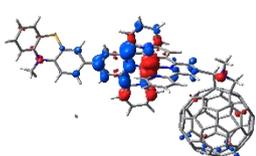 |
| $S_{49}$ | 3.77 | 329 | 0.3816 | 0.000 | 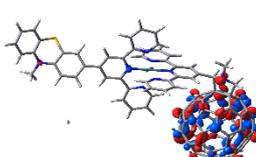 |

**Table S11.** The electronic couplings ($V_{DA}$) obtained by the generalized Mulliken–Hush (GMH) method as well as the fragment charge difference (FCD) approach for five **RuR** complexes.

| $V_{DA}$ (eV) | **RuH** | **RuPh** | **RuTol** | **RuAn** | **RuC$_{60}$** |
|---|---|---|---|---|---|
| GMH | 4.5x10⁻⁴ | 5.3x10⁻⁴ | 8.4x10⁻⁴ | 1.14x10⁻³ | 1.09x10⁻³ |
| FCD | 4.5x10⁻⁴ | 5.3x10⁻⁴ | 8.4x10⁻⁴ | 1.14x10⁻³ | 1.09x10⁻³ |

## Tamm-Dancoff approximation (TDA)

The Tamm-Dancoff approximation (TDA) often provides an improved description of the energy levels of triplet states.[1,2] However, as shown in a very recent theoretical study for structurally related (polypyridyl)-Ru(II)-based coordination compounds as well as in the scope of the present contribution, almost identical energetics are determined for triplet states at the TDDFT vs. the TDA levels of theory.[3] As summarized in Table S12 for the five complexes at hand, all driving forces are approximately -0.15 eV more favorable using TDA, while the reorganization energies are mostly unaffected. Furthermore, TDA-based electronic couplings are of the same magnitude as the respective TDDFT results. Therefore, a significant impact of the treatment of deexcitations can be excluded. The ET rates calculated at the TDA level of theory are roughly two orders of magnitude faster than the rates obtained using TDDFT as the driving forces are ~0.15 eV more favorable in TDA. However, the same trends are observed, *i.e.*, the decreasing rate from **RuH** to **RuPh**, **RuTol** and **RuAn**, and a faster rate for **RuC$_{60}$**.

**Table S12.** Driving forces ($\Delta G$), reorganization energies ($\lambda_D$, $\lambda_A$ and $\lambda_{AVG}$), electronic couplings ($V_{DA}$), rate constants ($k$) and lifetimes ($1/k$) for five **RuR** complexes as obtained at the TDA level of theory using the B3LYP hybrid functional.

|  | $\Delta G$ (eV) | $\lambda_i$ (eV) | $V_{DA}$ (eV) | $k_i$ (s$^{-1}$) |
|---|---|---|---|---|
| **RuH** | -0.2150 | 0.3175 | 10.0x10$^{-4}$ | 2.20x10$^{10}$ |
|  |  |  |  | 2.01x10$^{10}$ |
|  |  | 0.3291 |  | 2.10x10$^{10}$ |
|  |  | 0.3233 |  |  |
| **RuPh** | -0.1221 | 0.3561 | 8.0x10$^{-4}$ | 3.63x10$^9$ |
|  |  |  |  | 3.91x10$^9$ |
|  |  | 0.3493 |  | 3.76x10$^9$ |
|  |  | 0.3527 |  |  |
| **RuTol** | -0.0971 | 0.3546 | 10.5x10$^{-4}$ | 4.38x10$^9$ |
|  |  |  |  | 4.08x10$^9$ |
|  |  | 0.3610 |  | 4.23x10$^9$ |
|  |  | 0.3578 |  |  |
| **RuAn** | -0.0439 | 0.3564 | 12.5x10$^{-4}$ | 2.43x10$^9$ |
|  |  | 0.3679 |  | 2.12x10$^9$ |
|  |  | 0.3622 |  | 2.27x10$^9$ |
| **RuC$_{60}$** | -0.5843 | 0.4271 | 12.5x10$^{-4}$ | 2.27x10$^{10}$ |
|  | $\Delta G$ (eV) | 0.4938 |  |  |
|  |  | 0.4604 |  |  |